\documentstyle[12pt,hpatex,epsf]{article}
\input amssym.def
\input amssym.tex
       \def\eqref#1{(\ref{#1})}
\newcommand{\calC}{{\cal{C}}}
\newcommand{\calD}{{\cal{D}}}
\newcommand{\calE}{{\cal{E}}}
\newcommand{\calF}{{\cal{F}}}

\newcommand{\calH}{{\cal{H}}}
\newcommand{\calI}{{\cal{I}}}

\newcommand{\calQ}{{\cal{Q}}}
\newcommand{\calR}{{\cal{R}}}
\newcommand{\calS}{{\cal{S}}}

\newcommand{\sech}{\mbox{sech}\,}

\begin{document}
\title{Chaos in the Hill System}
\authors{C. Chicone} 
\address{Department of
Mathematics, University of Missouri,\\
Columbia, MO 65211}
\authors{B.  Mashhoon}
\address{Department of  Physics and
Astronomy, University of Missouri,\\ 
Columbia, MO 65211}  
\authors{and D. G.  Retzloff}
\address{Department of Chemical Engineering, University of Missouri,\\
Columbia, MO 65211}
\abstract{ 
We define the general Hill system and briefly 
analyze its dynamical behavior. A particular Hill
system representing the interaction of a Keplerian binary
system with a normally incident circularly polarized gravitational
wave is discussed in detail.
In this case, we compute the Poincar\'e-Melnikov function explicitly
and determine its zeros. Moreover, we provide numerical
evidence in favor of chaos in this system.
The partially averaged equations for the Hill system are used
to predict the regular behavior of the Keplerian orbit at resonance
with the external radiation.}
\section{Introduction} 
In a previous paper on gravitational ionization~\cite{5}, 
we discussed --- among other things --- the long-term
dynamical evolution of a Keplerian binary system
perturbed by a normally incident {\em circularly polarized}
gravitational wave. If the external radiation is approximated by
a monochromatic plane wave, then the dynamical system representing
the relative motion of the binary is very similar to the well-known
Hill system of celestial mechanics. Moreover, in our numerical
work on such a system, there appeared some preliminary evidence in
favor of Hamiltonian chaos (see figure~1 of~\cite{5}). The
purpose of the present work is to extend the notion of a Hill
system to include the external perturbation caused by a gravitational
wave. The generalized Hill system ---  which we refer to simply as
``the Hill system'' due to its close resemblance to the one 
originally introduced by Hill in his researches on the lunar
theory --- is presented in section~\ref{sec:2}. For the sake
of simplicity, we then choose a 
particular case for detailed analysis and discuss its dynamics
near resonance in section~\ref{sec:3}. Moreover, in section~\ref{sec:4}
the relevant Poincar\'e-Melnikov function is computed in this case
and its countable infinity of zeros are studied.
The possible relevance of this infinity of simple zeros
to the existence of chaos
in the Hill system is discussed. Numerical evidence for such chaos
is presented in section~\ref{sec:5} and a specific chaotic
orbit is described in some detail. 
In section~\ref{sec:v} we turn our attention
to the physics of such orbits in the inertial
frame in connection with the possible astrophysical relevance of our 
results. Section~\ref{sec:vii} is a discussion. 
Calculational details are relegated to the appendices.
\section{The Hill System}\label{sec:2}
Consider a Keplerian
binary system under the influence of an
external gravitational perturbation. In
fact, no binary system in the universe is
totally isolated as a consequence of the
universality of the gravitational
interaction. The binary is generally
affected by other masses as well as
gravitational radiation. We are interested
in a particular form of this interaction
that corresponds to Hill's celebrated
contribution to the theory of the motion of
the Moon~\cite{1}.

Let us therefore assume that the predominant
effect of the external perturbation is a
linear tidal force in the equation of {\em
relative} motion, so that this equation can
be written as
\begin{equation}\label{eq1}
\frac{d^2X^i}{dt^2}+\frac{k_0X^i}{r^3}=-
K_{ij}(t)X^j,
\end{equation}  where
$r=|{\bf X}|$, $k_0$ is a positive constant
and
$i,j=1,2,3$. Here the tidal matrix
$K(t)$ is in general symmetric and
traceless. Moreover, we assume that
$K_{13}=K_{23}=0$; indeed, a tidal matrix of
this form is consistent with our further
requirement that the perturbed relative
motion be confined to the $(X,Y)$-plane. To
arrive at the Hill system, we need to
restrict the form of
$K(t)$ even further; to this end, let us
suppose the following general form for
$K$:
\begin{eqnarray}\label{eq2}  
K_{11}&=&\xi _0+\xi _+\cos \Omega t-\xi _-\sin \Omega t,\nonumber \\
\nonumber K_{12} &=& \xi _+\sin \Omega t+\xi
_- \cos \Omega t,\\
\nonumber K_{22}&=&\xi_0-\xi_+\cos
\Omega t+\xi _-\sin \Omega t,\\
K_{33} &=&-2\xi _0, 
\end{eqnarray} 
where $\Omega$,  $\xi _0$ and $\xi_\pm$
are real constants.

It is possible to transform the equation of
relative motion \eqref{eq1} to a rotating
coordinate system via
$X^i=S_{ij}x^j$, where $S$ is an orthogonal
matrix given by 
\begin{equation}\label{eq6} 
S=\left[
\addtolength{\arraycolsep}{2\arraycolsep}
\begin{array}{ccc}
\cos\big(\frac{1}{2} \Omega t\big) & 
     -\sin \big(\frac{1}{2} \Omega t\big)
& 0\\*[.05in]
\sin \big(\frac{1}{2} \Omega t\big) & 
     \cos \big( \frac{1}{2}\Omega t\big)
& 0\\*[.05in]   
0 & 0 & 1
\end{array}
\right]
\end{equation}  
that represents a rotation
by an angle
$\frac{1}{2}\Omega t$ about the
$Z$-axis. The equation of relative motion
with respect to the rotating frame of
reference with coordinates
${\bf x}=(x,y,z)$ is then given by the
autonomous {\em Hill system}
\begin{eqnarray}\label{eq7}
\nonumber\frac{d^2x}{dt^2}-\Omega \frac{dy}{dt}-\frac{1}{4}
\Omega^2x+\frac{k_0x}{r^3}
&=&-(k_{11}x+ k_{12}y),\\
\label{eq8} \frac{d^2y}{dt^2}+\Omega
\frac{dx}{dt}-\frac{1}{4}\Omega^2
y+\frac{k_0y}{r^3} & =&-(k_{21}x+k_{22}y),\end{eqnarray} and
$z=0$, where $r=|{\bf x}|$ and
$k=S^{-1}KS$ is a {\em constant} matrix
given by $k_{11}=\xi_0+\xi _+$,
$k_{12}=\xi _-$, $k_{22}=\xi _0-\xi_+$,
$k_{13} =k_{23}=0$ and $k_{33}=-2\xi_0$.  It
is important to note that the sign of
$\Omega$  has not been specified; in fact,
$\Omega$ can be positive or negative. In
Hill's original system,
$\xi_0=-\Omega  ^2/8$, $\xi_+=-3\Omega ^2/8$
and $\xi _-=0$, corresponding to the
circular motion of the Earth-Moon system
about the Sun with frequency
$\Omega /2$. Hill's variational orbit, which
is a periodic solution of the Hill system,
has played a major role in the lunar
theory~\cite{2,3,4,guts}. In addition to 
the original system envisaged by Hill, the
general case includes a Keplerian binary
system that is tidally perturbed by a
normally incident circularly polarized
gravitational wave \cite{5}.

The Hill system may be obtained from the
Hamiltonian $H(p_x,p_y,x,y)$,
\begin{equation}\label{eq9}
H=\frac{1}{2}p^2-\frac{k_0}{r}-
\frac{1}{2}\Omega L_z+\frac{1}{2}
k_{ij}x^ix^j,\end{equation} where the
canonical momenta are given by
$p_x=\dot{x}-\Omega y/2$, $p_y=\dot{y}
+\Omega x/2$ and
$L_z=xp_y-yp_x$. Here $\dot{x}=dx/dt$ and
$L_z$ is the component of relative orbital
angular momentum normal to the orbital
plane. It is interesting to introduce in
this plane polar coordinates $(r,\theta )$
such that
$x=r\cos \theta$ and $y=r\sin \theta $. Then
with respect to the canonical variables
$(p_r,p_\theta ,r,\theta )$, equation
\eqref{eq9} can be written as
\begin{equation}\label{e10}
H=\frac{1}{2}\Big(p^2_r+
\frac{p^2_\theta}{r^2}\Big)
  -\frac{k_0}{r}- \frac{1}{2}\Omega p_\theta 
   +\frac{1}{2}r^2\Big(\xi _0 +\xi _+\cos
2\theta 
  +\xi _-\sin 2\theta \Big)
,\end{equation}  where $p_r={\bf x}\cdot
{\bf p} /r$ and $p_\theta =L_z$. In the
absence of the external perturbation, i.e.
when
$\Omega$, $\xi_0$ and $\xi_\pm$ vanish, the
relative motion of the binary can be
described in terms of a Keplerian ellipse.
In the presence of the perturbation, the
relative motion in the rotating system of
reference can be described at each instant
of time in terms of an {\em osculating
ellipse}. That is, if the perturbation is
turned off at an instant of time $t$, the
subsequent motion is along the osculating
ellipse at $t$. The energy
$E<0$ and orbital angular momentum
$p_\theta> 0$ of the osculating ellipse fix
its semimajor axis $a=-k_0/(2E)$ and
eccentricity $e$,  $0\leq e<1$, $p^2_\theta
=a(1-e^2)$. The position on the osculating
ellipse at time $t$ is measured from the
periastron by the true anomaly $v$, while
the orientation of this whole osculating
ellipse is determined in the orbital plane
by an angle $g$ such that $\theta =v +g$.
The equation of the osculating ellipse is
then given by $r=p_\theta ^2/(1+e\cos v
)=a(1-e\cos u)$, where $u$ is the eccentric
anomaly. Moreover,
$p_rp_\theta =e\sin v$, $(1-e\cos u)\sin v
=(1-e^2)^{\frac{1}{2}}\sin u$ and the mean
anomaly $\ell $ is given by
$\ell =u-e\sin u$. Only positive square
roots are considered throughout this paper.
The parameters of the osculating ellipse are
closely related to Delaunay's action-angle
variables. In fact, the Delaunay elements are
defined by $L=(k_0a)^{\frac{1}{2}}$,
$G=p_\theta $, $\ell$ and $g$ for noncircular orbits. 
It is proved
in \cite{4} that the change of variables
$(p_r,p_\theta ,r,\theta )\to (L,G,\ell, g)$
is canonical. In terms of these planar
Delaunay elements, we then have the
Hamiltonian for the Hill system in the form
\begin{equation}\label{eq11} H(L,G,\ell,g)
=-\frac{k_0^2}{2L^2}-\frac{1}{2}\Omega
G+\frac{1}{2}(\xi _0
\calQ+\xi_+\calC +\xi_-\calS).\end{equation}
Here
$\calQ=r^2$ is given by
\begin{equation}\label{eq12} \calQ
(L,G,\ell, g)=a^2\left[
1+\frac{3}{2}e^2-4\sum_{\nu =1}^\infty
\frac{\cos \nu \ell}{\nu^2} J_\nu (\nu
e)\right],\end{equation} and
$\calC =r^2\cos 2\theta $ and
$\calS =r^2\sin 2\theta$ are given in
Delaunay's elements by
\begin{equation}\label{eq13}
\calC+i\calS =a^2\exp(2ig)\left[
\frac{5}{2}e^2+\sum_{\nu =1}^\infty
\frac{1}{\nu} 
\Big( K^\nu _+\exp(i\nu\ell)+K^\nu_-\exp(-i\nu \ell)\Big)\right],
\end{equation}  
where
\[K^\nu_{\pm}=\frac{1}{2}\nu (A_\nu \pm B_\nu )\]  
and
\begin{eqnarray}\label{eq14}
A_\nu(e)&=&\frac{4}{\nu ^2e^2} [2\nu
e(1-e^2)J'_\nu(\nu e)-(2-e^2)J_\nu (\nu
e)],\\
\label{eq15}
B_\nu(e)&=&
-\frac{8}{\nu^2e^2}(1-e^2)^{\frac{1}{2}}
[eJ'_\nu (\nu e)-\nu (1-e^2)J_\nu (\nu e)].
\end{eqnarray}  
It is interesting to note
that $A_\nu =A_{-\nu}$ and $B_\nu
=-B_{-\nu}$, so that $K_\pm^{-\nu}=-K^\nu_\mp$.

It follows from our previous work
\cite{5} that the Kolmogorov-Arnold-Moser
(``KAM'') theory is applicable to the general Hill
system and that for sufficiently small
$\xi_0$ and
$\xi_\pm$ the motion is always bounded.
Moreover, our previous work (cf. figure~1 in~\cite{5}) 
indicates the presence of
Hamiltonian chaos under certain
circumstances. The purpose of the present
paper is to investigate further the nature
of this chaotic behavior. 

To simplify the analysis, we will assume in
the following that $\xi_0=0$ and that
$\xi_+=\epsilon \alpha \Omega ^2\cos
\varphi _0$ and $\xi_-=\epsilon \alpha
\Omega ^2\sin \varphi_0$. For $\Omega >0$,
this choice of parameters corresponds to the
tidal perturbation produced by an incident
right circularly polarized (i.e. positive helicity)
plane monochromatic gravitational wave of
amplitude $\alpha \epsilon$, frequency
$\Omega$ and phase constant $\varphi_0$
propagating along the $z$-axis. On the other
hand, we can let $\Omega \to -\Omega$ in
equations (2)--(7) and the resulting system
with the same choice of parameters would
correspond to left circularly polarized
(i.e. negative helicity) radiation of frequency $\Omega >0$; in this
way, our analysis covers both cases of
circular polarization. The {\em transverse}
nature of gravitational radiation implies
that for this case of normal incidence the
binary motion remains planar. The deviation
of the curved spacetime metric in the presence
of gravitational waves from the flat
Minkowski metric is characterized by the
perturbation parameter
$\epsilon$,
$0<\epsilon \ll 1$.  Efforts are under way
in several laboratories to detect
gravitational waves from astrophysical
sources with an expected amplitude of
$\epsilon \sim 10^{-20}$. Moreover, the
circular polarization amplitude $\alpha$ is
such that
$|\alpha |\sim 1$. To simplify matters even
further, we shall set $\varphi_0=0$. The
resulting Hill system has been discussed at
length in our previous
papers~\cite{5,cmr4,6}.  More generally, we
have investigated the long-term nonlinear
evolution of a Keplerian binary system
perturbed by external long-wavelength
gravitational waves as well as internal
gravitational radiation
damping~\cite{cmr4,6,7,75}. In fact,
gravitational {\em ionization} provided the
original motivation for our analysis
\cite{8}. The term ``ionization'' is derived
from a Greek word meaning ``to go'';
therefore, the concept of ionization need
not be restricted to electrically charged
systems such as in atomic physics. We have
shown that in the case under consideration
ionization does not in fact occur for $\epsilon<\epsilon_{\mbox{kam}}$. 
This is an
important consequence of the KAM theory and
implies, on the physical side, that the circularly polarized
gravitational wave does not steadily deposit
energy into the orbit. Indeed, in the
interaction of gravitational waves with a
binary system the energy in general flows
back and forth between the waves and the
binary.
Further discussion of gravitational ionization is contained
in section~\ref{sec:v} in connection with certain large-scale chaotic
behavior of the binary orbit for $\epsilon \gtrsim \epsilon_{\mbox{ch}}$,
where $\epsilon_{\mbox{ch}}$ corresponds to Chirikov's resonance-overlap
condition.

\section{Isoenergetic Reduction and Dynamics
Near Resonance}\label{sec:3}
With the simplifications already mentioned, 
the equations of motion
derived from the Hamiltonian
\eqref{eq11} are given by
\begin{eqnarray} 
\dot{L}&=&-\frac{1}{2}\epsilon\alpha
\Omega^2 
 \frac{\partial \calC}{\partial
\ell},\nonumber\\
\dot{G} &=& -\frac{1}{2}\epsilon\alpha
\Omega ^2\frac{\partial \calC}{\partial
g},\nonumber\\
\dot{\ell } & =& \frac{k_0^2}{L^3}
+\frac{1}{2}\epsilon\alpha \Omega ^2
\frac{\partial \calC}{\partial L},
\nonumber\\
\dot{g} & = &-\frac{1}{2} \Omega
+\frac{1}{2}\epsilon\alpha \Omega^2
 \frac{\partial \calC}{\partial G}.\label{eq16}
\end{eqnarray}  
These equations provide a
classic example of a perturbation problem in
mechanics. The unperturbed system is
expressed in action-angle variables and is
completely integrable. We are interested in
the dynamics of the perturbed system for
small $\epsilon$. This is the ``fundamental
problem in dynamical systems'', according to
Poincar\'e~\cite{2}.

System \eqref{eq16} is a 2-degree-of-freedom autonomous
Hamiltonian system. Thus by a well-known
method that we will now describe, it can be
reduced to a
$1\frac{1}{2}$-degree-of-freedom  system
once the total energy of the system is
fixed. Let us describe the reduction
technique for the general case. In fact, we
will consider the system
\begin{eqnarray}\label{eq17} 
\dot{q} &=&\frac{\partial \hat{H}}{\partial
p}(q,p,\vartheta ,I), \nonumber\\ 
\dot{\vartheta } &
=&\frac{\partial\hat{H}}{\partial I}
(q,p,\vartheta ,I),\nonumber\\
\dot{p} & =&-\frac{\partial
\hat{H}}{\partial q}(q,p,\vartheta ,I),
\nonumber\\ 
\dot{I}&=&-\frac{\partial \hat{H}}{\partial
\vartheta}(q,p,\vartheta ,I),
\end{eqnarray}  where the Hamiltonian
function $\hat{H}$ is assumed to be periodic
in the angular variable
$\vartheta$ and there is some region $\calR$
of the phase space in which the function
$\partial \hat{H}/\partial I$ does not
vanish.  Under these assumptions, we can
solve for $I$ as a function of the remaining
variables on an energy surface, 
$$\{ (q,p,\vartheta
,I):\hat{H}(q,p,\vartheta ,I)=h\},$$  to
obtain 
$$I={\hat K}(q,p,\vartheta ,h).$$  Moreover,
because $\partial \hat{H}/\partial I$ does
not vanish in
$\calR$, the variable $\vartheta$ is either
increasing or decreasing along all orbits in
$\calR$. Thus,
$\vartheta$ behaves like a temporal variable
and  can be taken to be a new independent
variable.  Indeed, the system
\begin{equation}\label{eq18}
\frac{dq}{d\vartheta} =
\frac{\partial \hat{H}}{\partial
p}\big/\frac{\partial \hat{H}}{\partial
I},\quad \frac{dp}{d\vartheta
}=-\frac{\partial \hat{H}}{\partial
q}\big/\frac{\partial \hat{H}}{\partial
I}\end{equation} is not singular in
$\calR$.\ If $\vartheta \mapsto ({\hat
u}(\vartheta) ,{\hat v}(\vartheta))$  is a
solution of the system~\eqref{eq18}, then
$$
\dot{\vartheta }=
\frac{\partial \hat{H}}{\partial I}({\hat
u}(\vartheta) ,{\hat v}(\vartheta),
\vartheta, {\hat K}({\hat u}(\vartheta
),{\hat v}(\vartheta ),\vartheta ,h)).  
$$  The solution $t\mapsto \vartheta (t)$ of
this scalar differential equation can be
used to obtain a solution of the original
system
\eqref{eq17}; namely,
$$ q(t) ={\hat u}(\vartheta (t)),\quad p(t)
={\hat v}(\vartheta (t)),
\quad I(t)={\hat K}(q(t),p(t),\vartheta
(t),h).
$$

To obtain a simpler form for the
system~\eqref{eq18}, let us use the equation
$$\hat{H}(q,p,\vartheta , {\hat
K}(q,p,\vartheta, h))=h$$  to obtain the
identities
$$\frac{\partial \hat{H}}{\partial
q}+\frac{\partial \hat{H}}{\partial I}
\frac{\partial {\hat K}}{\partial q}=0,\quad
\frac{\partial \hat{H}}{\partial
p}+\frac{\partial \hat{H}}{\partial
I}\frac{\partial {\hat K}}{\partial p}=0,$$
and, in turn, system \eqref{eq18} in the form
\begin{equation}\label{eq19}
\frac{dq}{d\vartheta}=-
\frac{\partial\hat{K}}{\partial
p}(q,p,\vartheta ,h),
\quad 
\frac{dp}{d\vartheta }=\frac{\partial
\hat{K}}{\partial q}(q,p,\vartheta ,h).
\end{equation} Under our assumptions, the
system
\eqref{eq19} is periodic in its independent
variable. Moreover, system
\eqref{eq19} is in the form of a
``time-dependent''
$1\frac{1}{2}$-degree-of-freedom Hamiltonian
system with Hamiltonian ${-\hat
K}(q,p,\vartheta ,h)$. Let us now employ
this reduction principle in the study of the
system~\eqref{eq16}.

For the Hamiltonian \eqref{eq11}, we recall
that if $\epsilon$ is sufficiently small,
then the KAM Theorem applies and all orbits
in a region $\calR$ of an energy surface
remain bounded. In particular, if $\calR$ is
a closed subset of  a bounded region, then
the function $\partial \calC / \partial G$
is  bounded over $\calR$. Thus, if
$\epsilon$ is sufficiently small, then
$\dot{g}<0$ along all orbits starting in
$\calR$. Also, by the Implicit Function
Theorem, if the constant energy surface is
given by
$$H(L,G,\ell,g)=h,$$ then there is an
implicit solution $G=F(L,\ell,g)$ such that
$$H(L,F(L,\ell,g),\ell,g)=h.$$ Using the
reduction principle given above, the reduced
system in our case is given by
\begin{equation}\label{eq20}
\frac{d\ell}{dg}=-\frac{\partial F}{\partial
L}(L,\ell,g),\quad
\frac{dL}{dg}=\frac{\partial F}{\partial
\ell}(L,\ell,g).\end{equation}

To solve for $G$ in the equation
$H(L,G,\ell,g)=h$, let us suppose that in
the corresponding equation~\eqref{eq11} we
have
$G=G_0+\epsilon G_1+O(\epsilon^2)$ and then
equate coefficients to obtain the solution up
to first order in the small parameter. 
Let us recall here that the orientation of the background
inertial coordinate system is so chosen that $G_0>0$ by convention.
Our
computations yield the values
$$ G_0 =-\frac{2}{\Omega}\Big(h_0+
\frac{k_0^2}{2L^2}\Big),\quad  G_1 =\alpha
\Omega\,
\calC(L,G_0,\ell,g)-\frac{2}{\Omega}h_1,
$$  where $h_0$ and $h_1$ are constants such
that $h=h_0+\epsilon h_1$; that is, $h_0$ is
the unperturbed energy in the rotating frame.
Thus, we have that
\begin{eqnarray}
\frac{d\ell}{dg}&=&-\frac{2k_0^2}{\Omega
L^3}-\epsilon \alpha \Omega
\Big(\frac{\partial \calC}{\partial
L}(L,G_0,\ell, g) +\frac{2k_0^2}{\Omega
L^3}\frac{\partial \calC}{\partial G}
(L,G_0,\ell,
g)\Big)+O(\epsilon^2),\nonumber\\
\frac{dL}{dg}&=&\epsilon \alpha \Omega
\frac{\partial \calC}{\partial
\ell}(L,G_0,\ell ,g)+O(\epsilon
^2).\label{eq21}\end{eqnarray}  Also, let us
note that in view of equation~\eqref{eq13}
the $1\frac{1}{2}$-degree-of-freedom system
\eqref{eq21} is periodic with period $\pi$
relative to its independent variable.

The unperturbed system \eqref{eq21} is {\em
a priori} stable; that is, it has no
hyperbolic orbits. In particular, there are
no homoclinic or heteroclinic orbits that
can be used to locate transverse homoclinic
or heteroclinic points, and, in turn,
chaotic invariant sets, after perturbation.
Rather, if such transverse homoclinic or
heteroclinic points exist in the perturbed
system, they must be created directly from
the perturbation. As is well known, this
fact precludes a direct application of the
usual ``Melnikov'' approach to determine the
existence of transverse homoclinic or
heteroclinic points. The problem is that the
separatrix splitting distance
is easily computed to be proportional to $\epsilon
M+O(\epsilon ^2)$, where $M$ is the Melnikov
integral. In the usual case,
$M$ is independent of $\epsilon$ and thus
the first term of the indicated series is
the dominant term. However, in the {\em a
priori} stable case, $M$ depends on
$\epsilon$ and is in fact exponentially
small. Therefore, it is not clear if
$\epsilon M(\epsilon )$ is the dominant term
in the expression for the splitting distance.
On the other hand, some rigorous results do
exist for systems similar to those that are
encountered in the Hill system. Consider,
for instance, the rapidly forced pendulum
given in the form
$$\ddot{\phi }+\sin \phi =\epsilon^p\sin
(\Omega t/\epsilon),$$  where $\epsilon $ is
a small parameter and $p$ is a positive
integer. If $p>3$, then the Melnikov
integral is exponentially small, but the
first-order term is still dominant, see~\cite{ds}
and ~\cite{eks}.

While the rigorous analysis of the
separatrix splitting problem remains unclear
at this writing, we will carry through the
first-order analysis for Hill's system. Once
the rigorous analysis is settled at some
future date, we hope that the formulation
reported here will prove to be valuable.

To proceed with the perturbation theory for
Hill's system, we will use a standard
technique: we will ``blow up'' a resonance
and partially average the resulting
equations to produce a system, obtained by a
change of coordinates, that has 
homoclinic orbits. We will then compute the
Melnikov integral for this system.
For a regular perturbation problem where the Melnikov function
does not depend on the perturbation parameter, 
the simple zeros of this function would
indicate the presence of chaos in the system.
After a blow up at a resonance and a reparametrization of the
system to slow time, the Melnikov function does depend
on the perturbation parameter. It is for this reason that we are
not able to make a rigorous argument that the existence of
simple zeros implies that the system is chaotic.

For system \eqref{eq21} that is periodic
in $g$ with period $\pi$, resonance would occur
when this period is commensurate with the period
of ``fast'' motion in $\ell$, i.e. $\pi\Omega L^3/k_0^2$. 
Therefore, the resonances are
given by relations of the form
$$m\frac{k_0^2}{L^3}=n\Omega$$ where $m$ and
$n$ are relatively prime integers. This
resonance condition fixes the magnitude of
$L$; therefore, let $L_0$ denote the
resonant value of $L$ at the $(m:n)$
resonance. To blow up the resonance, we will
use the change of coordinates given
by
$$
L=L_0+\epsilon^{1/2} \rho,\quad 
\ell =-\frac{2n}{m} g+\eta. 
$$ 
System~\eqref{eq21}, in the new coordinates,
has the form
\begin{eqnarray}
\frac{d\rho}{dg}&=&\epsilon^{1/2} \alpha
\Omega \calC_\ell  +\epsilon \alpha \Omega
\rho (\calC_{\ell L}
 +\frac{2n}{m}\calC_{\ell G})+O(\epsilon ^{3/2}),\nonumber\\
\frac{d\eta}{dg}&=&\epsilon^{1/2}
\frac{6n}{m L_0}\rho
 -\epsilon \Big(\frac{12n}{m L_0^2}\rho^2+\alpha
\Omega (\calC_L+\frac{2n}{m}\calC_G)\Big) 
+O(\epsilon ^{3/2}),\label{eq22}
\end{eqnarray} 
where the function $\calC$ and
its partial derivatives are evaluated
at
\[( L_0,G_0(L_0),\eta-2ng/m ,g).\] For
notational convenience, let us define
$\mu=\epsilon^{1/2}$ and express the second-order 
approximation of system \eqref{eq22} in the form
\begin{eqnarray} 
\frac{d\rho}{dg} & =& \mu\gamma (\eta ,g) 
  +\mu^2 \rho \kappa (\eta ,g ),\nonumber\\
\frac{d\eta }{dg} & =&\mu c \rho
  -\mu^2\beta (\rho,\eta,g),\label{eq23}
\end{eqnarray}  where $c$ is a constant
given by $c=6n/(mL_0)$ and
\begin{eqnarray*}
\gamma(\eta,g)&=&
\alpha\Omega\calC_\ell,\qquad
\kappa(\eta,g)= \alpha\Omega\big(\calC_{\ell
L}+\frac{2n}{m}\calC_{\ell G}\big),\qquad
\beta(\rho,\eta,g) = \frac{2c}{L_0}\rho^2 
+\alpha\Omega\big(\calC_L+\frac{2n}{m}
\calC_G\big).
\end{eqnarray*} 
This system is in the
correct form for temporal averaging. To this
end, we define $\langle \gamma \rangle$ to
be the average of $\gamma$ over the periodic
temporal variable $g$ with period $m\pi$ and
$\lambda$ to be the deviation of $\gamma$
from its mean,
\begin{eqnarray} 
\langle \gamma \rangle (\eta ): & =&
\frac{1}{m\pi }
  \int^{m\pi}_0\gamma (\eta
,g)\,dg,\nonumber\\
\lambda (\eta,g): &=& \gamma
(\eta,g)-\langle \gamma \rangle (\eta).
   \label{eq24}
\end{eqnarray}  Moreover, let
$\Lambda$ denote the solution of the partial
differential equation
$\partial \Lambda/\partial g=\lambda$  with
the side condition that its average should
vanish; that is,
$\langle \Lambda \rangle (\eta )=0$. Using
the averaging transformation
$$\rho=\bar\rho+\mu \Lambda
(\bar\eta,g),\quad \eta =\bar\eta,$$  system
\eqref{eq23} takes the form
\begin{eqnarray} 
\frac{d\bar\rho}{dg} & =&\mu\langle \gamma
\rangle (\eta)
  +\mu^2\bar\rho
\big(\kappa(\eta,g)-c\frac{\partial
\Lambda}{\partial\eta}\big)
    +O(\mu^3),\nonumber\\
\frac{d\eta }{dg}&=&\mu c\bar\rho+\mu^2
 (c\Lambda (\eta,g)
-\beta(\bar\rho,\eta,g))+O(\mu^3).
\label{eq25}
\end{eqnarray} It is useful to introduce
dimensionless quantities $J$ and $\Gamma$
such  that
$$\bar\rho=L_0 J,\quad \Lambda =L_0\Gamma;$$
then, the second-order approximation of
system \eqref{eq25} has the convenient form
\begin{eqnarray}  
\frac{dJ}{dg} & =&\mu
\frac{1}{L_0}\langle \gamma \rangle (\eta
)+\mu^2 J
\Big(\kappa (\eta ,g)
    -\frac{6n}{m}\frac{\partial
\Gamma}{\partial\eta}\Big),\nonumber\\
\frac{d\eta }{dg} & =& 6\frac{n}{m}\mu
J+\mu^2
\Big(6\frac{n}{m} \Gamma (\eta ,g)-\beta
(L_0J,\eta ,g)\Big).\label{eq26}
\end{eqnarray}

Recall formula~\eqref{eq13} and the fact that
$$
\gamma =
\alpha
\Omega\,\calC_\ell\Big(L_0,G_0(L_0),-
\frac{2n}{m} g+\eta,g\Big),
$$  and note that $\langle
\gamma \rangle (\eta )$ is the real part of
\begin{eqnarray*}
\lefteqn{\frac{1}{m\pi}\int^{m\pi}_0\alpha \Omega 
  \Big(\calC_\ell\big(L_0,G_0(L_0),-\frac{2n}{m}g+\eta,g\big)
  +i\calS_\ell \big(L_0,G_0(L_0),-\frac{2n}{m} g+\eta ,g\big)\Big)\,dg} \\  
&=& i\frac{\alpha \Omega a^2}{m\pi}\int^{m\pi}_0 \exp(2ig)
\sum^\infty_{\nu=1} 
\Big(K^\nu_+\exp(i\nu \eta-i\frac{2n}{m}\nu g)-
   K^\nu_-\exp(-i\nu \eta+i\frac{2n}{m}\nu g)\Big)\, dg.
\end{eqnarray*}  
This integral vanishes
unless $n=1$ and
$m=\nu$, in which case its value is
$i\alpha \Omega a^2K^m_+\exp(im\eta)$. 
Using the fact that
$a=L^2/k_0$, $L=L_0$ at resonance and 
$\Omega =mk_0^2L^{-3}_0$, the average of
$\gamma$  at the $(m:1)$ resonance can be written in the form
\begin{equation}\label{eq27}
\frac{1}{L_0}\langle \gamma \rangle (\eta
)=-\alpha mK^m_+\sin m\eta.
\end{equation} If in system \eqref{eq26} we
substitute using formula \eqref{eq27},
change to the new angular variable $\sigma
:=m\eta -\pi$ and the slow temporal
variable $\tau :=6\mu g$, then we obtain the
equivalent system
\begin{eqnarray} 
\label{eq28}\frac{d\sigma }{d\tau } & =&
J+\mu
\left[\Gamma \Big(\frac{\sigma
+\pi}{m},\frac{\tau }{6\mu}\Big)
-\frac{m}{6}\beta\Big(L_0J,
\frac{\sigma+\pi}{m},
   \frac{\tau }{6\mu}\Big)\right],  \\*[.05in]
\frac{dJ}{d\tau } &=& \frac{\alpha m }{6}
K^m_+ \sin\sigma 
  +\mu J\left[\frac{1}{6}\kappa
\Big(\frac{\sigma +\pi }{m},
\frac{\tau}{6\mu}\Big) -\frac{\partial
\Gamma}{\partial\sigma}
\Big(\frac{\sigma +\pi }{m},
\frac{\tau}{6\mu}\Big)\right].
\label{eq28a}
\end{eqnarray}

\begin{figure}[ht]
\epsfxsize=300pt
\epsfysize=300pt
\centerline{\epsffile{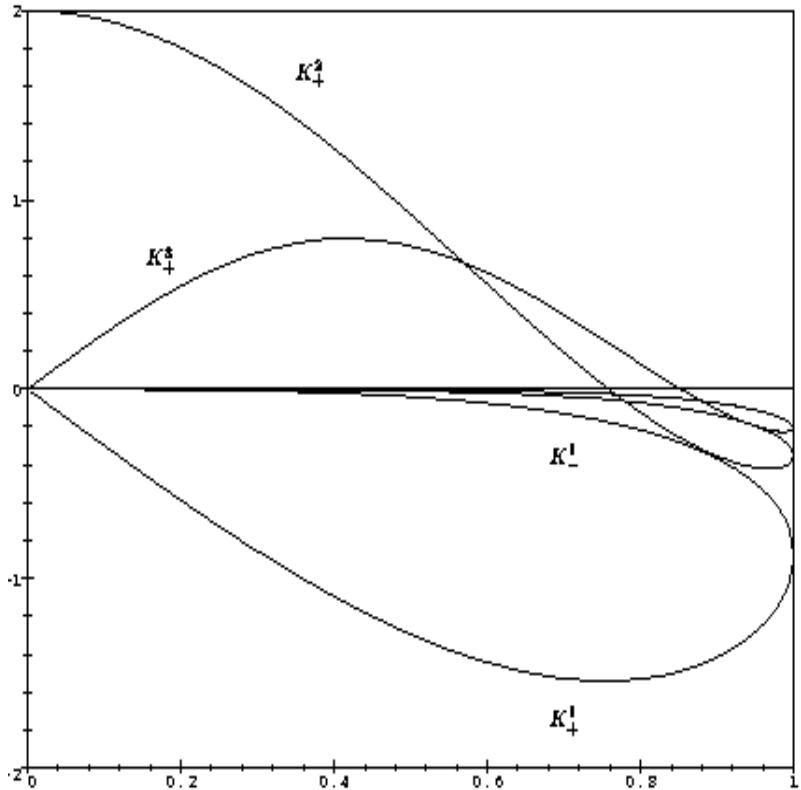}} 
\caption[]{ 
Plot of $K^m_\pm$ for $m=1,2,3$. Note that
$K^m_+$ and $K^m_-$ meet at $e=1$. Some of the properties of
$K^m_\pm$ are discussed in Appendix~\ref{appen:a}.
}
\end{figure}            
Let us note that the unperturbed form of the 
system \eqref{eq28}--\eqref{eq28a},
given by 
\begin{equation}\label{eq:un}
\frac{d\sigma_0}{d\tau}=J_0,\qquad
\frac{dJ_0}{d\tau}=-\delta\sin\sigma_0,
\end{equation} 
is the phase-plane system for a mathematical pendulum. 
Here
\[\delta:= -\frac{\alpha m}{6} K^m_+, \]
$m=1,2,3,\ldots$,
is the order of the resonance, the constant $\alpha$ could be positive
or negative and $K^m_+$, $2K^m_+=m(A_m+B_m)$ as defined by 
equations~\eqref{eq14} and~\eqref{eq15}, 
is a function of the eccentricity at resonance
$(e_0)$. 
Some properties of $K^m_\pm(e)$ near $e=0$ and $e=1$ are
given in Appendix~\ref{appen:a}.
A graph of the functions $K_\pm^m$ for $m=1,2,3$ 
is given in figure~1, which illustrates the analytical results
of Appendix~\ref{appen:a}. 
Our explicit calculations refer to the
case of incident positive helicity radiation and
hence only $K^m_+$ is involved in the definition
of $\delta$ (see section~\ref{sec:v}).  
It follows from inspection of figure~1 that
depending upon the value of the eccentricity $e_0$, $\delta$ could be
positive, negative or zero.
If $\delta>0$, then the phase portrait has hyperbolic
saddle points at $(J_0,\sigma_0) =(0,\pm \pi)$ that are
connected by the heteroclinic solutions
\begin{eqnarray*}
\sigma_0(\tau)&=&2\arcsin(\tanh(\delta^{1/2}\tau))
                     =2\arctan(\sinh(\delta^{1/2}\tau)),\\
J_0(\tau)&=&\pm 2 \delta^{1/2}\sech(\delta^{1/2}\tau).
\end{eqnarray*}
If $\delta<0$, then 
the phase portrait has hyperbolic
saddle points at $(J_0,\sigma_0) =(0,0)$ and  $(J_0,\sigma_0) =(0,2\pi)$  
that are
connected by the heteroclinic solutions
\begin{eqnarray*}
\sigma_0(\tau)&=&2\arcsin(\tanh(|\delta|^{1/2}\tau))+\pi
                     = 4\arctan (\exp(|\delta|^{1/2}\tau)),\\
J_0(\tau)&=&\pm 2 |\delta|^{1/2}\sech(|\delta|^{1/2}\tau).
\end{eqnarray*}                                           
All of these heteroclinic orbits are also
heteroclinic manifolds for the corresponding unperturbed stroboscopic
Poincar\'e map with period
$12\pi \mu$. If $\delta=0$, then the
Poincar\'e-Melnikov approach is not directly applicable; hence, we need to
exclude this possibility in the calculation of the Melnikov function.
It is therefore clear from the results
of Appendix~\ref{appen:a} that all resonant circular orbits
$(e_0=0)$ are excluded from our analysis except for $m=2$. This is 
particularly significant in view of a result of {\em linear}
perturbation analysis~\cite{8} that the main resonance
for a circular orbit occurs for $m=2$; this is consistent with
the reciprocity between the emission and absorption of gravitational
radiation. Furthermore,
we need to exclude $e_0\approx 0.76$ for the
$(2:1)$ resonance,  $e_0\approx 0.85$ for the
$(3:1)$ resonance, etc., since $\delta=0$ at these zeros
of $K^m_+(e)$.
Let us remark here that the initial orbit is such that $p_\theta>0$
by convention; therefore, there is a marked difference between the
absorption of positive and negative helicity gravitational waves. 
In particular, $K^m_-(e)$ does not vanish for $0<e<1$.
Let us also note here that $\sigma_0$ is an
odd function of $\tau$ while $J_0$ is an even function; 
moreover, $J_0\to 0$ as 
$\tau\to\pm\infty$.  As mentioned previously, we
will compute the Melnikov integral along
these invariant manifolds. For regular
perturbations, the existence of simple zeros
for this function implies that, for sufficiently small
$\mu$, there are chaotic invariant sets for
the perturbed Poincar\'e map, and hence
there are chaotic invariant sets for the
perturbed system. In the system
\eqref{eq28}--\eqref{eq28a}, the
perturbation is not regular due to the rapid
oscillations of the perturbation when $\mu$
is small. Nevertheless, we will compute the
Melnikov integral for this system.

\section{Poincar\'e-Melnikov Function}\label{sec:4}
Recall that for a time-periodic planar
system of the form
$$\dot{x} =f_0(x,y)+\mu f_1(x,y,t),\quad
\dot{y}=g_0(x,y)+\mu g_1(x,y,t),$$  where
the unperturbed system is Hamiltonian, if
$t\mapsto (x(t),y(t))$ is a solution of the
unperturbed system that starts on a
heteroclinic or homoclinic orbit, then the
Melnikov function defined on this orbit is
given by
\begin{equation}\label{eq29} 
M(\xi)=\int^\infty_{-\infty}
\big[f_0(x(t),y(t))g_1(x(t),y(t),t+\xi)\\ -
g_0(x(t),y(t))f_1(x(t), y(t),
t+\xi)\big]\,dt.
\end{equation}

The computation of $M$ for the system
\eqref{eq28}--\eqref{eq28a} is quite
complex. However, the end result is
reasonably simple. Let us begin with the
term $\Gamma =\Lambda /L_0$;  as in the
computation of formula~\eqref{eq27}, we have
that 
\begin{equation}\label{eq30}
\frac{\gamma }{L_0}=-\alpha
m\sum^\infty_{\nu =1} \left[K^\nu _+\sin
\Big(\nu \eta +2g
(1-\frac{\nu }{m})\Big) +K^\nu
_-\sin \Big( \nu \eta -2g
( 1+\frac{\nu }{m})\Big)\right].\end{equation}
Using formulas \eqref{eq27} and
\eqref{eq30}, let us compute
\[
\frac{\Lambda}{L_0}=\frac{1}{L_0}\int^g
\lambda\,dg
=\frac{1}{L_0}\int^g(\gamma-\langle
\gamma\rangle)\,dg
\] such that $\langle \Lambda
\rangle(\eta)=0$; in fact, the constant of
integration must vanish and the result is 
\begin{equation}\label{eq:27}
 \Gamma=\frac{1}{2}\alpha m\left[\,
\sum^\infty_{\stackrel{\nu=1}{\scriptscriptstyle
\nu\neq m}} K^\nu _+\frac{\cos \Big(\nu
\eta +2g ( 1-\frac{\nu }{m})\Big)}{1-\frac{\nu }{m}}
 -\sum^\infty_{\nu =1}K_-^\nu
\frac{\cos \Big(\nu \eta -2g ( 1+\frac{\nu }{m})\Big)}
{1+\frac{\nu}{m}}\right].
\end{equation} 
Let us recall that in the definition of the Poincar\'e-Melnikov
function~\eqref{eq29},
$$ f_0=J_0,\quad f_1=\Gamma-\frac{m}{6}\beta
,\quad g_0=-\delta\sin \sigma_0 ,\quad  
g_1=J_0\big(\frac{1}{6}
\kappa-\frac{\partial\Gamma}{\partial
\sigma}\big),$$ 
where
$$\beta =\frac{12}{m} J_0^2+\alpha \Omega
(\calC_L+\frac{2}{m}\calC_G),\quad 
\kappa=\alpha
\Omega ( \calC_{\ell L}+\frac{2}{m}\calC_{\ell G}).$$ 
Thus, we have a preliminary expression for the integrand of
$M$ (cf. Appendix~\ref{appen:b}) 
\begin{equation}\label{29}
f_0g_1-g_0f_1=\frac{\alpha \Omega}{6}J_0^2
(\calC_{\ell L}+\frac{2}{m}
\calC_{\ell G})
-J_0^2\frac{\partial\Gamma}{\partial \sigma}
 +\delta\sin \sigma_0 f_1.
\end{equation}

The Melnikov integral \eqref{eq29} is
calculated in Appendix~\ref{appen:b} and the result is 
\begin{eqnarray}
M(\xi ) & = &
\frac{5}{6}\alpha m^2 H_m I^0 \calS^0 \nonumber\\ 
&&\quad {}+\frac{1}{2}\alpha m^3\sum^\infty_{\nu =1}
  \Big(\frac{K^\nu_+}{\nu -m} P_\nu^+\calS^+_\nu
 -\frac{K^\nu_-}{\nu +m} P^-_\nu \calS^-_\nu\Big)\nonumber\\
&&\quad {}+\frac{7}{6}\alpha
m^2\sum^\infty_{\nu =1} (K^\nu_+ P^+_\nu
\calS^+_\nu +K^\nu_-P^-_\nu \calS^-_\nu)\nonumber\\
&&\quad {}+\frac{1}{6} \alpha m^2 F_m 
   \sum^\infty _{\nu =1}(K^\nu _{+e}\,P^+_\nu \calS^+_\nu 
     +K^\nu_{-e}\,P^-_\nu \calS^-_\nu).\label{30}
\end{eqnarray}  
Here the only quantities that depend on
$\xi $ are $\calS^0 =\sin (\xi /3\mu )$ and
\[\calS^\pm _\nu =\sin [\pm (\xi /3\mu )(1\mp \nu /m)+\nu \pi /m],\]
while $H_m$, $F_m$,
$K^\nu_\pm$ and $K^\nu_{\pm e}=dK^\nu_\pm
/de$ all depend upon the eccentricity of the
orbit at resonance
$e_0=(1-\hat{e}_0^2)^{\frac{1}{2}}$, where
$\hat{e}_0=G_0/L_0$. 
In fact,
$F_m=(\hat{e}_0^2-2\hat{e}_0/m)/e_0$ and
$H_m=e_0(2e_0+F_m)$. Moreover, $I^0$
and $P^\pm_\nu$ are also dependent upon $e_0$ via $\delta$ and are
given by
\begin{eqnarray}
I^0 & = & \delta
\int^\infty_{-\infty}\sin \Big( \frac{\tau }{3\mu}\Big) 
\sin \sigma_0(\tau )\,d\tau,\nonumber\\ 
P^\pm_\nu & = & {}\pm \frac{1}{3\mu \nu}\Big( 1\mp
\frac{\nu }{m}\Big)
\int^\infty_{-\infty}J_0(\tau )\cos \left[
\frac{\nu}{m}\sigma _0(\tau )\pm \frac{\tau
}{3\mu}\Big( 1\mp
\frac{\nu}{m}\Big)\right]\, d\tau.\label{31}
\end{eqnarray}
It is interesting to note that $\mu I^0$ and $\mu P_\nu^\pm$ only depend
upon $\mu'=\mu|\delta|^{1/2}$.
A detailed discussion of these integrals is contained in Appendix~\ref{appen:b};
in fact, a method is described there that can be used to calculate
$I^0$ and $P_\nu^\pm$ whenever $2\nu/m$ is an integer.
For instance, it follows from the results of Appendix~\ref{appen:b} that
$P_\nu^{\pm}$ can be explicitly determined for $m=1,2$.

It is important to point out that the
Poincar\'e-Melnikov function is periodic
with period $6\pi \mu m$ and has a countable
infinity of zeros at $\xi_N =3\mu (1+mN)\pi$
for any integer $N$. Indeed, $\calS^0$ and
$\calS^\pm_\nu$ all vanish for $\xi_N /(3\mu)=(1+mN)\pi$, where
$N=0,\pm 1,\pm 2,\ldots $; therefore,
$M(\xi_N)=0$. 
 
The zeros of the Melnikov function $M(\xi)$ are all generally expected
to be simple.
This may be seen from the nature of two consecutive zeros of
$M(\xi)$, e.g.\ $\xi_0$ and $\xi_1$, due to the periodicity
of the Melnikov integral. However, it is more convenient for our 
purposes to compute $M'(\xi_N)$ for general $N$. We note that
in the expression~\eqref{30} for $M(\xi)$ only
$\calS^0$ and $\calS_\nu^\pm$ depend upon $\xi$, and 
\begin{eqnarray*}
\frac{d\calS^0}{d\xi}(\xi_N)&=&-\frac{1}{3\mu}(-1)^{Nm},\\
\frac{d\calS^\pm_\nu}{d\xi}(\xi_N)&=&
  {}\mp\frac{1}{3\mu}(-1)^{N(m-\nu)}(1\mp\frac{\nu}{m}).
\end{eqnarray*}
It follows that
\begin{eqnarray}
-(-1)^{Nm}\, 3\mu M'(\xi_N)&=&\frac{5}{6}\alpha m^2 H_m I^0
+\frac{2}{3}\alpha m^2\sum_{\nu=1}^\infty (-1)^{N\nu}
  (K^{\nu}_{+} P_{\nu}^+ -K^{\nu}_-P_\nu^-)\nonumber\\
&& {}-\frac{7}{6}\alpha m\sum_{\nu=1}^\infty(-1)^{N\nu}
  \nu(K^\nu_+ P_\nu^+ +K^\nu_- P_\nu^-)\nonumber\\ 
&& {}\label{n:33}+\frac{1}{6}\alpha m^2 F_m\sum_{\nu=1}^\infty(-1)^{N\nu}
  \big[(1-\frac{\nu}{m}) K^\nu_{+e} P_\nu^+
                 -(1+\frac{\nu}{m})K^\nu_{-e}P_\nu^-\big],
\end{eqnarray}
assuming that term-by-term differentiation of the infinite
series in~\eqref{30} is meaningful. 

Inspection of equation~\eqref{n:33} indicates that in general
$M'(\xi_N)\ne 0$. To see this in some detail, let us focus
attention on the value of $M'(\xi_N)$ for an eccentricity
$e_0$ such that $0\le e_0\ll 1$.  
We remark that $\mu^2M'(\xi_N)$ can be regarded as a function
of independent variables $e_0$ and $\mu'=\mu |\delta|^{1/2}$; in
fact, $\mu'$ occurs only in $I^0$ and $P_\nu^\pm$.
The forms of $H_m(e)$ and $F_m(e)$ then
lead us to distinguish two cases:
$m\ne2$ and $m=2$. Suppose first that $m\ne 2$; then,
the discussion following the introduction of $\delta\ne 0$
in equation~\eqref{eq:un} implies that $e_0\ne 0$. It
follows from the
results of Appendix~\ref{appen:a} 
that the leading term in the expansion of $M'(\xi_N)$ in powers of
$e_0\ll 1$ is 
\[\frac{(-1)^{N(m+1)}(m-2)\alpha}{6m\mu e_0}[(m-1)P^+_1-(m-3)P^+_3],\]
where $P_\nu^+$ is given by equation~\eqref{31}.
The general form~\eqref{b:5} given in
Appendix~\ref{appen:b} for the integrals in equation~\eqref{31}
can be used to show that
this leading term is nonzero for $m=1$; a similar result is expected for $m> 2$.
Let us next suppose that $m=2$. If $e_0=0$, then
the Melnikov function $M(\xi)$ vanishes in this case; further analysis
of this limiting case is beyond the scope of this paper. For 
$e_0\ne 0$, the results of Appendix~\ref{appen:a} 
imply that the leading term in the expansion of $M'(\xi_N)$ in powers
of $e_0\ll 1$ is now
\[\frac{(-1)^{N}\alpha e_0}{6\mu}(P^+_1+25 P^+_3),\]
where $P_1^+$ and $P_3^+$ are given by equation~\eqref{31} for $m=2$.
The general form~\eqref{b:5} given in Appendix~\ref{appen:b} for the
integrals in equation~\eqref{31} again indicates that
this leading term is nonzero.
Hence, $M(\xi)$ has in general simple zeros for $0<e_0\ll 1$.
\section{Numerical Experiments}\label{sec:5}
We now consider the same Hamiltonian as in
the calculation of the Poincar\'e-Melnikov
function,
\begin{equation}\label{eq32}
H=\frac{1}{2}\Big(
p^2_r+\frac{p^2_\theta}{r^2}\Big)-\frac{k_0}{r}
   -\frac{1}{2}\Omega p_\theta 
   +\frac{1}{2}\epsilon \alpha \Omega^2 r^2\cos 2\theta,
\end{equation} 
and
will demonstrate numerically that this
system could be chaotic near a resonance. We
first fix an energy surface $H(p_r,p_\theta, r,\theta )=h$ 
and consider all orbits
confined to this energy surface. For
instance, choosing an orbit with 
$(p_r,p_\theta ,r, \theta)=(e,1,1,0)$ at $t=0$ fixes
\[
h=\frac{1}{2}(1+e^2)-k_0-\frac{\Omega
}{2}+\frac{1}{2}\epsilon\alpha \Omega ^2.
\] 
The equations of motion are
\begin{eqnarray}\frac{dr}{dt}&=&p_r,\nonumber\\
\frac{d\theta }{dt}&=&\frac{p_\theta
}{r^2}-\frac{\Omega}{2},\nonumber\\
\frac{dp_r}{dt} & = &
\frac{p_\theta^2}{r^3}-\frac{k_0}{r^2}-\epsilon
\alpha \Omega^2 r\cos 2\theta,\nonumber\\
\frac{dp_\theta}{dt}&=&\epsilon \alpha \Omega^2 r^2\sin 2\theta,\label{33}
\end{eqnarray} 
which can
be integrated with different initial
conditions all with
$H(p_r,p_\theta ,r,\theta )=h$. This energy
equation can be used to calculate $p_\theta$
algebraically in terms of the other
variables; we limit our attention here to
the solutions of this quadratic equation
with $p_\theta >0$. Moreover, we choose
$\theta$ as our independent variable and
note that the resulting system for
$(p_r,r)$ is periodic in $\theta$ with
period $\pi$. We then consider the
Poincar\'e map and plot in the
$(p_r,r)$-plane the result of our numerical
integration of the system at $\theta =0,\pi,2\pi,\ldots$. 
In following this procedure, we encounter the difficulty
that $\theta$ is not necessarily monotonic.
Despite this complication, it is possible
to obtain useful information from the integration of system~\eqref{33}
as in our previous work (see figure~1 of~\cite{5}).
To avoid this problem altogether, let us restrict attention to
those orbits that stay with a simple branch of $p_\theta>0$. That is,
we note that $H(p_r,p_\theta,r,\theta)=h$ has the solution
\[
p_\theta=\frac{1}{2}\Omega r^2\big[ 1\pm(1-U)^{1/2}\big],
\]
where
\[
U=\frac{8}{\Omega^2 r^2}(\frac{1}{2}p_r^2-\frac{k_0}{r}-h)
+4\epsilon\alpha\cos 2\theta.
\]
Thus the system~\eqref{33} reduces to 
\begin{eqnarray}
\frac{dr}{dt}&=&p_r,\nonumber \\
\frac{d\theta}{dt}&=&\pm \frac{1}{2}\Omega(1-U)^{1/2},\nonumber \\
\frac{d p_r}{dt}&=&-\frac{1}{r}\, p_r^2+\frac{k_0}{r^2}+\frac{2h}{r}
 +\frac{1}{2}\Omega^2 r
   \big[ 1-4\epsilon \alpha \cos 2\theta \pm(1-U)^{1/2}\big],
\label{n:35}
\end{eqnarray}
where the upper (lower) sign corresponds to the positive (negative) branch
of $\dot\theta$. We have subjected the system~\eqref{n:35},
{\em restricted to the positive branch of} $\dot\theta$, 
to numerical analysis and have
found numerical evidence for Hamiltonian chaos as presented in 
figure~2.

It is interesting to consider scale transformations in the dynamical
systems~\eqref{33} and~\eqref{n:35} along the lines described in detail
in our previous work~\cite{6,7,75}. The result is that we
can use dimensionless quantities and set $k_0=1$ once we use 
the initial semimajor axis of the binary system
as our unit of length 
and the initial period of the binary divided by
$2\pi$ as our unit of time. 
We shall use this convention in the numerical experiments
reported in this paper.

\begin{figure}[p]
\epsfxsize=250pt
\epsfysize=250pt
\centerline{\epsffile{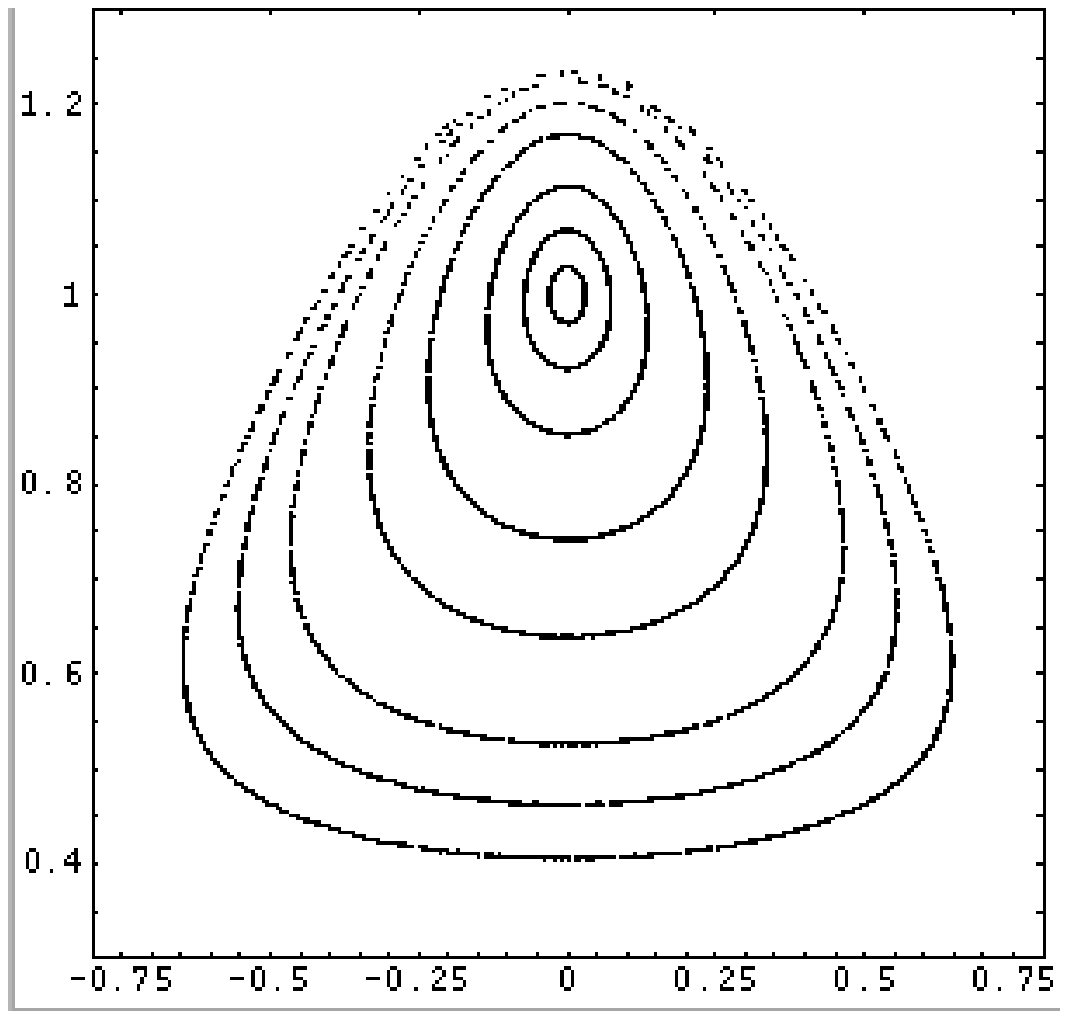}}
\epsfxsize=250pt
\epsfysize=250pt
\vspace*{.25in}
\centerline{\epsffile{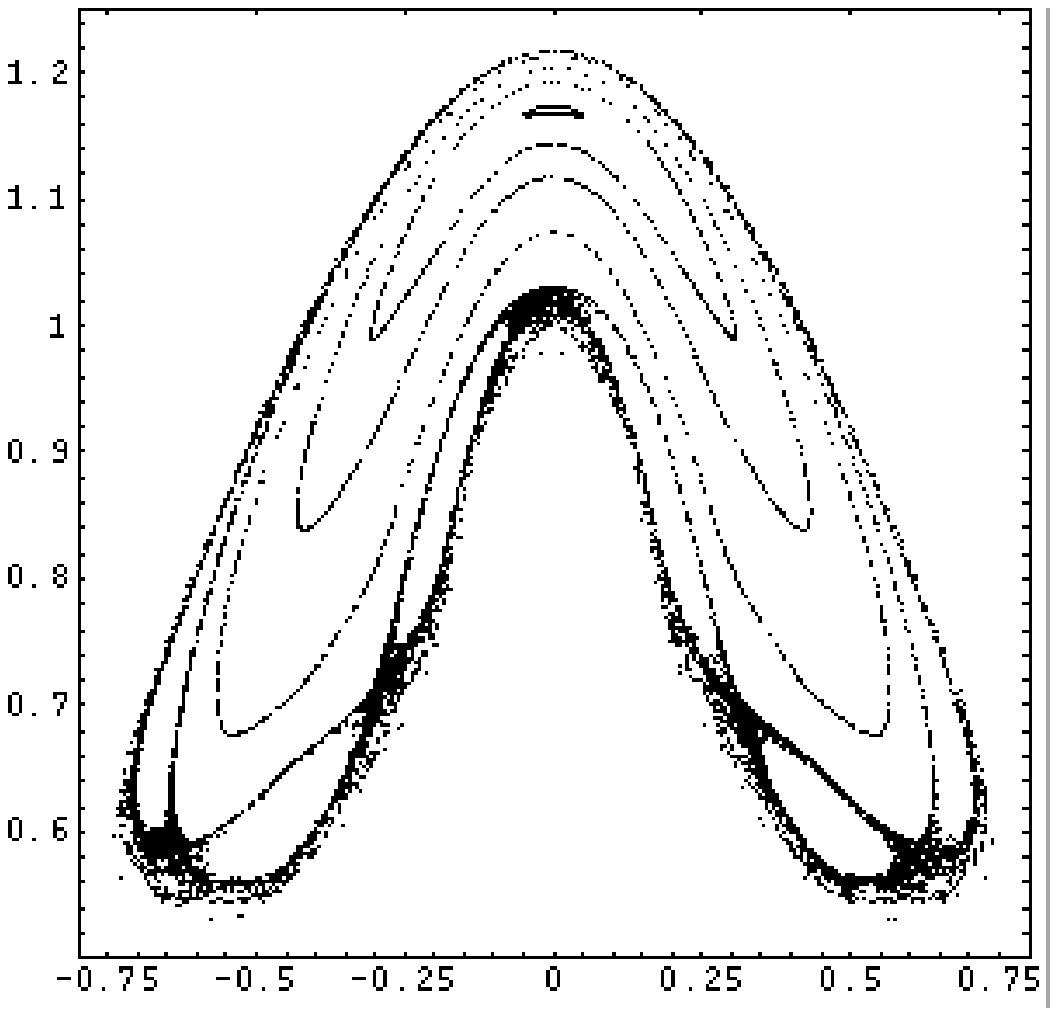}}
\caption[]{ The phase portrait of the Poincar\'e map for system~\eqref{n:35}
with the parameters given in displays~\eqref{figdata1}--\eqref{figdata3}.
The top panel is a phase portrait for the unperturbed system~\eqref{figdata2} 
while the bottom panel depicts a phase portrait for a perturbed 
system~\eqref{figdata3} on a nearby energy surface.
}
\end{figure} 
\begin{figure}[p]
\epsfxsize=300pt
\epsfysize=300pt
\centerline{\hspace{.4in}\epsffile{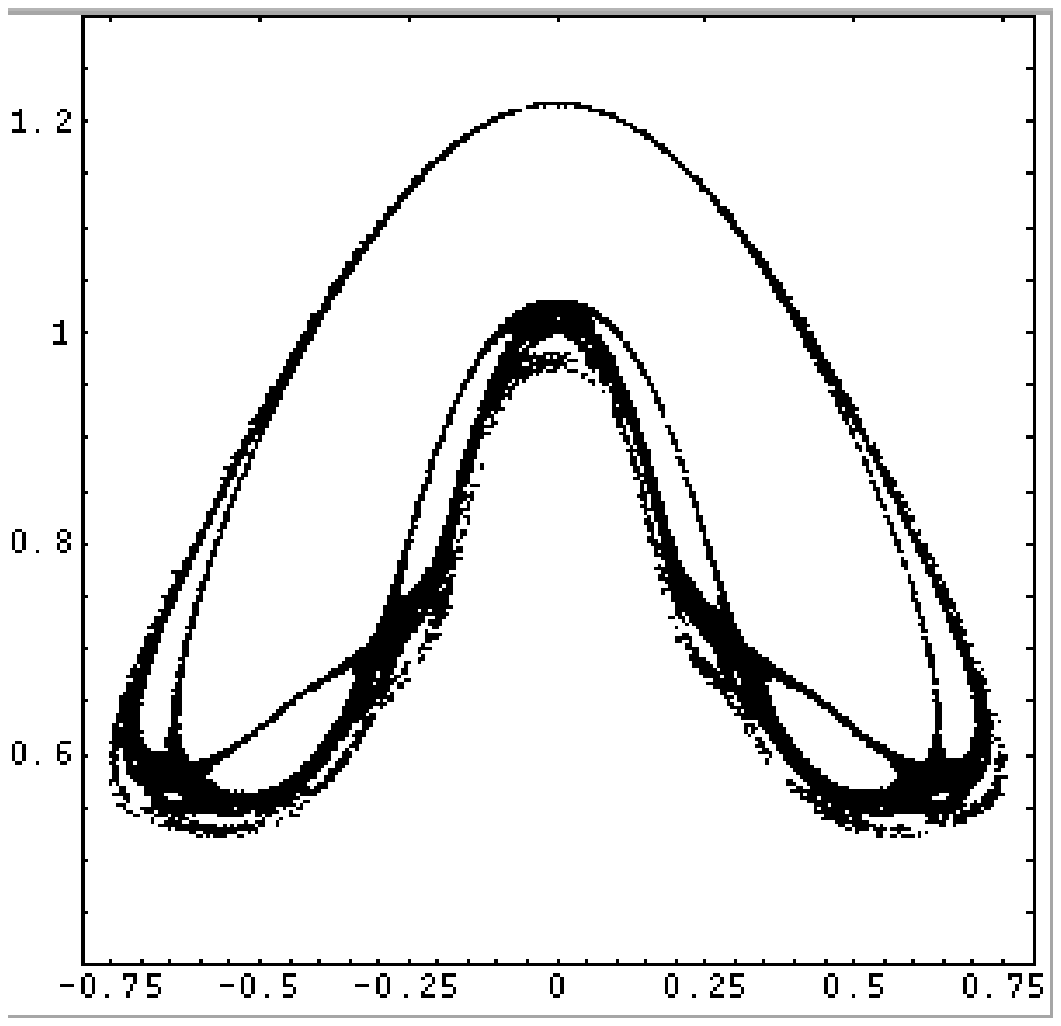}}
\epsfxsize=300pt
\vspace*{.25in}
\centerline{\hspace*{.25in}\epsffile{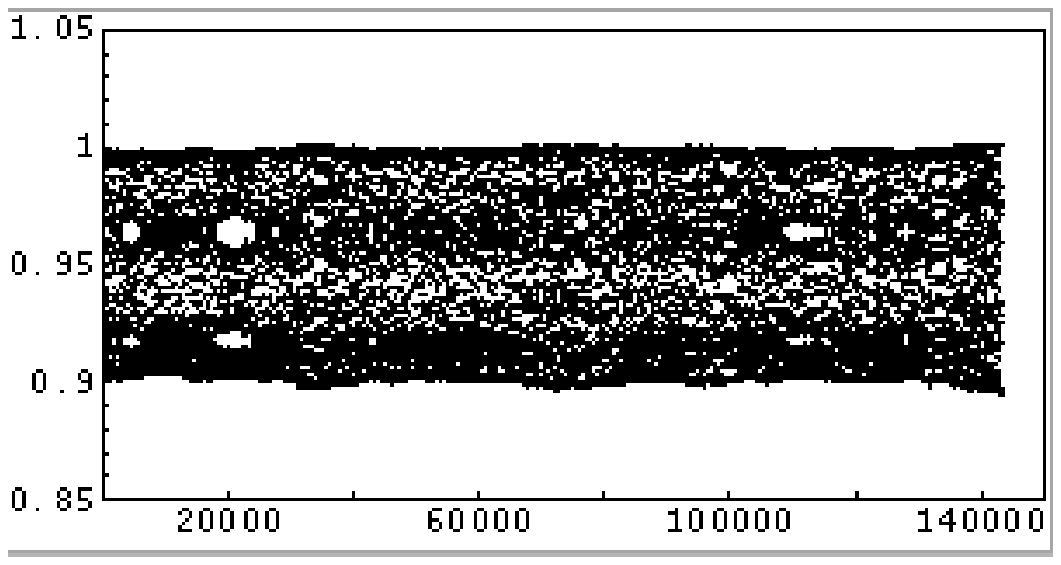}}
\caption[]{The top panel shows a single orbit of the
Poincar\'e map for system~\eqref{n:35}
with the parameters given in displays~\eqref{figdata1} and~\eqref{figdata3}.
The bottom panel is a graph of the Delaunay variable $L$ versus time for the
same orbit. It follows from this plot that the behavior of the
semimajor axis ($a=L^2$) of the osculating ellipse is
{\em chaotic} in this case.
}
\end{figure} 
Partial phase portraits of the Poincar\'e map for
system~\eqref{n:35} are depicted in figures~2 and~3, where we 
have chosen the parameter values ($k_0=1$)
\begin{equation}\label{figdata1}
\Omega =1,\qquad \alpha=2,
\end{equation}
for the plane right circularly polarized gravitational wave
that propagates perpendicularly to the orbital plane.
Also, for the top panel of figure~2 the external perturbation
is absent so that
\begin{equation}\label{figdata2}
\epsilon=0, \qquad h=-0.999982,
\end{equation} 
while for the bottom panel of figure~2 and  for figure~3
\begin{equation}\label{figdata3} 
\epsilon=0.0131, \qquad h=-0.986882.
\end{equation} 
In these graphs, the horizontal axis corresponds to the variable $p_r$ while
the vertical axis corresponds to $r$. 
Inspection of the system~\eqref{n:35} reveals that the phase 
portrait of the Poincar\'e map is in general symmetric about the
line $p_r=0$.

The top panel of figure~2 depicts several orbits corresponding
to invariant tori for the unperturbed system ($\epsilon=0$),
whereas the bottom panel depicts several orbits for the perturbed system
($\epsilon\ne 0$).
Let us note that the stochastic region in figure~2 is obtained
as a single orbit. This same orbit
is also shown in the top panel of figure~3.
It can be obtained via integration of system~\eqref{n:35}, or~\eqref{33},
with 
\[ (r,\theta)=(0.61927711963654,0)\]
and
\[ (p_r,p_\theta)=(0.51405620574951,0.83454334164660)\]
as the initial conditions and with the parameters given in
equations~\eqref{figdata1} and~\eqref{figdata3}.
This chaotic orbit is near
the unperturbed $(1:1)$ resonant torus. 
Let us recall here that the energy of the osculating ellipse
is given by
\begin{equation}\label{eq:ener}
E=\frac{1}{2}\big(p_r^2+\frac{p_\theta^2}{r^2}\big)-\frac{k_0}{r},
\end{equation}
so that the semimajor axis is
$a=-k_0/(2E)$ and the corresponding Delaunay element is
$L=(k_0 a)^{1/2}$. The $(m:n)$ resonance occurs at 
$mk_0^2/L_0^3=n\Omega$; however, in the case under
consideration here the primary resonances are $(m:1)$ resonances
and our chaotic orbit is near the dominant $(1:1)$ resonance. 
Thus  $L$ for the chaotic orbit is likely to fluctuate
about the $(1:1)$ resonant value $L_0=1$, though in the case under
consideration we expect that secondary $(m:n)$ resonances with $n> 1$
are also involved since $\epsilon$ is large enough.
The bottom panel
of figure~3 is a graph of the Delaunay variable $L$ versus time for
the same orbit depicted in the top panel. 
The stochastic region filled out by the orbit
depicted in figure~3, an orbit of a $2$-degree-of-freedom time-independent
Hamiltonian, seems to be bounded by KAM tori. 
The fine structure of the graph of $L$ versus $t$  over
short time scales, i.e.\  a few thousand units of time, shows both chaotic
and apparently regular oscillations. These are accounted for by reference
to the orbit of the corresponding Poincar\'e map shown in
the top panel of figure~3. The corresponding trajectory visits regions
near hyperbolic saddle points where transversal crossings of stable
and unstable manifolds induce a homoclinic tangle with embedded
Smale horseshoes, etc. Thus a sojourn near such a region produces
a highly chaotic apparent change in the semimajor axis of the
corresponding osculating ellipse. However, when the orbit is 
``distant'' from these hyperbolic saddle points, the semimajor axis
of the  osculating ellipse oscillates with apparent regularity though
the regimes of apparent regularity are dependent on the scale at
which the fluctuations in $L$ are observed. 
This results in a form of Hamiltonian intermittency.
Of course, the sojourn time for apparently chaotic (large-amplitude
chaotic signal) versus apparently regular (small-amplitude chaotic signal)
oscillations is itself highly chaotic in our numerical experiments. 
 
The width of the stochastic layer $(\delta L)$ in figure~3 may be estimated
on the basis of the following considerations: If $\epsilon=0$, one
finds that the variation of $E=h+\Omega p_\theta/2$ vanishes along
the orbit (cf. eq.~\eqref{33}) and hence $L$ is constant
$(\delta L/L=0)$. Physically, this comes about since
an ellipse is simply viewed from a rotating frame. Once
the perturbation is turned on, the KAM theory implies that for
sufficiently small $\epsilon>0$ the motion is bounded for all time.
Inspection of the system~\eqref{n:35} then reveals that the small
parameter in the problem is effectively $4|\alpha|\epsilon$ as
$0\le |\cos 2\theta| \le 1$; therefore, we expect that 
$\delta L/L\approx 4|\alpha|\epsilon$ in agreement with the
data of figure~3.

The phase portrait of the Poincar\'e map
for system~\eqref{n:35} exhibits extreme sensitivity to
changes in the parameters of the system. For example, stochastic
regions comparable in relative size to the one depicted in figure~3 are 
not easily detectable for a random choice of parameter values. 
We have verified all of the main numerical results reported in this
paper via two different stiff integration routines using double
precision arithmetic.
We emphasize that a detailed numerical study of the Hill
system is beyond the scope of this paper, since
the relevant parameter space is rather large.
Our numerical results nevertheless support the
view that there is chaos in the Hill system.
On the other hand, it would be interesting
from an astrophysical standpoint to see the
chaotic behavior of the binary in the
inertial frame of reference. This issue is
taken up in the next section.

\section{Inertial Observers and the Hill System}\label{sec:v}
The equation of motion~(1) can be derived
from the Hamiltonian
$$\calH=\frac{1}{2}P^2-\frac{k_0}{r}+
\frac{1}{2}K_{ij}X^iX^j$$ 
in the inertial frame of reference. This Hamiltonian 
can be written in polar coordinates
$(r,\Theta )$ as
\begin{equation} 
\calH = \frac{1}{2} \Big(
P^2_r+\frac{P^2_\Theta}{r^2}\Big)-
\frac{k_0}{r}+\frac{1}{2}r^2[\xi _0 +\xi_+\cos
(2\Theta -\Omega t)+\xi _-\sin (2\Theta -\Omega t)],\label{34}
\end{equation} 
where
we have used equation (2) for the tidal matrix. 
Let us recall here that $\Omega $ can be
positive or negative.

The connection between the inertial motion
and the Hill motion based on Hamiltonian~\eqref{e10}
can be simply obtained by referring the
inertial motion based on the Hamiltonian
\eqref{34} to the rotating frame. In this process,
$\Theta =\theta +\Omega t/2$ and the
canonical momenta remain invariant, i.e.
$P_r=p_r$ and $P_\Theta =p_\theta$, while
the Hamiltonian in the inertial frame is
related to that in the rotating frame by the
well-known relation
\begin{equation}\label{35}
\calH=H+\frac{1}{2}\Omega p_\theta\end{equation}
that expresses angular momentum-rotation
coupling. It is important to observe that
with the initial conditions  $(p_r,p_\theta ,r,\theta )$ 
at $t=0$ being common to both
the inertial and rotating frames, we can
integrate the equations of motion in these
frames starting from the same initial
osculating ellipse.

Let us now choose $\xi_0=\xi_-=0$ and
$\xi_+=\epsilon\alpha \Omega ^2$ as before. It then
follows that the perturbation in the
inertial Hamiltonian is of the form
$\frac{1}{2}\epsilon\alpha \Omega^2r^2\cos (2\Theta \mp\Omega t)$, 
where the upper (lower) sign
stands for plane right (left) circularly
polarized gravitational radiation of
frequency $\Omega >0$ that is normally
incident on the Keplerian binary system. To
describe the {\em inertial} motion, we introduce
the osculating ellipse with semimajor axis
$\tilde{a}$, eccentricity
$\tilde{e}$, etc., in the inertial frame and
the associated Delaunay elements
$(\tilde{L},\tilde{G},\tilde{\ell},\tilde{g})$.
The equations of motion in terms of
Delaunay's variables are 
\begin{eqnarray}\frac{d\tilde{L}}{dt}&=&\epsilon
\frac{\partial \calH_{\mbox{ext}}}{\partial
\tilde{\ell}},\nonumber\\
\frac{d\tilde{G}}{dt} & = & {}-\epsilon
\frac{\partial \calH_{\mbox{ext}}}{\partial
\tilde{g}},\nonumber\\
\frac{d\tilde{\ell}}{dt} & = &
\frac{k^2_0}{\tilde{L}^3}+\epsilon
\frac{\partial \calH_{\mbox{ext}}}{\partial
\tilde{L}},\nonumber\\
\frac{d\tilde{g}}{dt} & =& \epsilon
\frac{\partial \calH_{\mbox{ext}}}{\partial
\tilde{G}},\label{36}\end{eqnarray} 
and
$$\frac{d\tilde{\tau }}{dt}=\Omega,$$ where
$\tilde{\tau}=\Omega t$ is a new variable and
$$\calH_{\mbox{ext}}(\tilde{L},\tilde{G},
\tilde{\ell},\tilde{g},\tilde{\tau})=\frac{1}{2}
\alpha
\Omega ^2r^2\cos (2\Theta \mp
\tilde{\tau}).$$
We are interested in the general behavior of system~\eqref{36} taking
into consideration the fact that the motion is always bounded 
for $\epsilon<\epsilon_{\mbox{kam}}$ as a consequence
of the KAM Theorem.

There are two frequencies in the dynamical
system under consideration here, namely the
Keplerian frequency $\omega =k^2_0/\tilde{L}^3$ and
the frequency of the external radiation
$\Omega$; hence, the possibility of
resonance cannot be ignored. 
Thus the averaging principle is in general inapplicable here due to
the fact that resonance could occur.
Indeed, resonance could come about
when $\omega$ and
$\Omega$ are commensurate, i.e. relatively
prime integers $m$ and
$n$ exist such that $m\omega =n\Omega$; at 
this $(m:n)$ resonance,  $\tilde{L}$
is fixed at its resonant value
$\tilde{L}_0$. It will be shown in this section that for
system~\eqref{36} the {\em primary} resonances are in fact
$(m:1)$ resonances, i.e. $n=1$. The resonant behavior of the system
around the $(m:1)$ resonance manifold can then be described by
the partially averaged equations for system~\eqref{36}.  

It is important to point out here that the $(m:n)$ resonance
is {\em invariant}, i.e.\  the resonance is the same whether it is
encountered in the rotating frame of reference or in the inertial
frame of reference. To make this point explicit, let us note that
the energy of the osculating ellipse
\[
\calE=\frac{1}{2}\big(P_r^2+\frac{P_\Theta^2}{r^2}\big)-\frac{k_0}{r}
\]
is an invariant quantity in the sense described above ($\calE=E$), 
and hence so is
the semimajor axis of the resonant orbit and its corresponding
Delaunay element $\tilde L=L=(-k_0^2/2\calE)^{1/2}$ that 
becomes fixed at resonance. The same holds for the
angular momentum ($\tilde{G}=G$) and the eccentricity
($\tilde{e}=e$) of the {\em osculating} ellipse; 
nevertheless, the tildes on the Delaunay action variables will
be generally kept in the rest of this section as a reminder that
the Hill system is being viewed from the inertial reference frame.
Chaos is invariant as well and  expected to occur near a 
resonance; indeed, this point can be illustrated with a single
chaotic orbit from the case studied in figure~2. This chaotic orbit
that is near a $(1:1)$ resonance is given in figure~3.
Further illustration of small-amplitude chaos is contained
in figures~4 and~5 described below.
Figure~6 illustrates large-amplitude chaos that comes about when
primary resonances overlap. 

The average behavior of the
system around the $(m:1)$ resonance manifold can be
obtained from the second-order partially averaged
dynamics. To this end, let
$$\tilde{L}=\tilde{L}_0+\epsilon^{\frac{1}{2}}\calD,
\quad
\tilde{\ell}=\frac{k_0^2t}{\tilde{L}_0^3}+\Phi,$$
such that the second-order averaged
equations  {\em in the inertial frame} are given by~\cite{7,75}
\begin{eqnarray}
\dot{\calD} & =& -\epsilon^{\frac{1}{2}}
   (-mT_c\sin m\Phi +mT_s\cos m\Phi )\nonumber\\
&&\quad {}-\epsilon \calD
 (-m\frac{\partial T_c}{\partial \tilde{L}}\sin m\Phi
    + m\frac{\partial T_s}{\partial \tilde{L}}\cos m\phi ),\nonumber\\
\dot{\tilde{G}} & = & -\epsilon (\frac{\partial T_c}{\partial
\tilde{g}}\cos m\Phi +\frac{\partial
T_s}{\partial \tilde{g}}\sin m\Phi),\nonumber\\
\dot{\Phi} & = & 
-\epsilon^{\frac{1}{2}}\frac{3k^2_0\calD}{\tilde{L}^4_0}+\epsilon
(
\frac{6k^2_0\calD^2}{\tilde{L}_0^5}+\frac{\partial
T_c}{\partial
\tilde{L}}\cos m\Phi +\frac{\partial
T_s}{\partial \tilde{L}}\sin m\Phi
),\nonumber\\
\dot{\tilde{g}} & = & \epsilon (
\frac{\partial T_c}{\partial
\tilde{G}}\cos m \Phi 
+\frac{\partial T_s}{\partial \tilde{G}}\sin m\Phi).\label{37}
\end{eqnarray} 
Here
$T_c=\calF^m_\pm (\tilde{L},\tilde{G})\cos
2\tilde{g}$ and
$T_s=\mp
\calF^m_{\pm}(\tilde{L},\tilde{G})\sin
2\tilde{g},$ where
\begin{equation}\label{38} \calF^m_\pm
=\frac{\alpha}{2m}\Omega ^2
\tilde{a}^2K^m_\pm (\tilde e)\end{equation} for
$n=1$. If $n\neq 1$, $T_c=T_s=0$ and
primary resonances do not occur.
It follows that for system~\eqref{36} the primary
resonances are given by $m\omega=\Omega$.  The system~\eqref{37} is
evaluated at $\tilde{L}=\tilde{L}_0$.

The second-order averaged dynamics can be
simplified if we introduce a new variable
$\Delta =m\Phi \pm 2\tilde{g}$. It is then
straightforward to recast \eqref{37} into the form
\begin{eqnarray}\dot\calD & = & m\epsilon
^{\frac{1}{2}}({\calF}^m_\pm
+\epsilon^{\frac{1}{2}}\calD\calF^m_{\pm\tilde{L}})
\sin\Delta,\nonumber\\
\dot{\tilde{G}}& = & \pm 2\epsilon
\calF_\pm^m\sin
\Delta,\nonumber\\
\dot{\Delta} & = & -\epsilon
^{\frac{1}{2}}\frac{3\Omega}{\tilde{L}_0}\calD
+\epsilon \frac{6\Omega }{\tilde{L}^2_0}\calD^2+\epsilon
(m\calF^m_{\pm \tilde{L}}\pm
2\calF^m_{\pm \tilde{G}})\cos \Delta,\label{eq:39}\end{eqnarray} 
where $\calF^m_{\pm
\tilde{L}}=\partial \calF^m_\pm /\partial
\tilde{L}$, etc. Restricting attention to
the {\em first-order} averaged equations in
\eqref{eq:39}, we find that this system can be
integrated. 
Thus
\begin{equation}\label{40}\tilde{G}-\tilde{G}_0=\pm
\frac{2}{m}(\tilde{L}-\tilde{L}_0),
\end{equation}
where $\tilde{G}_0$ is the orbital angular
momentum when the system is exactly at
resonance. Moreover,
\begin{equation}\label{41} 
\calD^2=\frac{\alpha m}{3} \tilde{L}_0^2K^m_\pm (\tilde{e}_0)
(\cos\Delta -\cos \Delta _0),
\end{equation} 
where
$\Delta_0$ and $\tilde{e}_0$ belong to the
relative orbit at resonance. It follows from equation~\eqref{41}
that $\calD$ has oscillatory character
and the maximum value of
$\calD^2$ occurs at either 
$\Delta =(0,2\pi ,4\pi ,\ldots )$ or $(\pi ,3\pi, 5\pi ,\ldots )$ 
depending upon whether $\alpha K^m_\pm(\tilde{e}_0)$ 
is positive or negative, respectively.
The amplitude of the {\em total} variation in $L$ is thus
given by $\delta L=2\epsilon^{1/2}\calD_{\mbox{max}}$, where
\begin{equation}\label{eq:48}
\calD_{\mbox{max}}^2
  =\frac{2\alpha m}{3}\tilde{L}^2_0 K^m_\pm(\tilde{e}_0)
\sin^2\big(\frac{m}{2}\Phi_0\pm \tilde g_0 \big)
\end{equation}
for $\alpha K^m_\pm>0$, and 
\begin{equation}\label{eq:49}
\calD_{\mbox{max}}^2
  =-\frac{2\alpha m}{3}\tilde{L}^2_0 K^m_\pm(\tilde{e}_0)
\cos^2\big(\frac{m}{2}\Phi_0\pm \tilde g_0 \big)
\end{equation}
for $\alpha K^m_\pm<0$.
Other properties of the motion can be studied as well since
$\Delta$ varies just like a standard nonlinear pendulum.
In addition to these results from the first-order averaged
equations, small regular corrections exist that are beyond the
scope of this paper and are due to terms of second order in
$\epsilon^{1/2}$ in the averaged equations. Furthermore, certain
chaotic effects are also expected near the primary
$(m:1)$ resonance.

These theoretical conclusions may be illustrated by a simple
example: Imagine a resonant orbit
with $(p_r,p_\theta ,r,\theta)=(\tilde{e},1,1,0)$ at $t=0$ such that
$\tilde{e}=1/2$,
$\tilde{g}_0=-{\pi}/{2}$,
$\tilde{u}={\pi}/{3}$,
$\tilde{v}={\pi}/{2}$ and hence $\Phi_0 ={\pi}/{3} -3^{1/2}/4$ for this
orbit. We choose $k_0=1$, $\epsilon
=10^{-3}$, $\alpha =2$ and $\Omega
=m/\tilde{L}^3_0=m(3/4)^{3/2}$, so
that the orbit is
initially in $(m:1)$ resonance. The results of the
numerical integration of this system for
$m=1,2$ are in good agreement with the predictions of the averaged
equations~\eqref{40} and~\eqref{41}, since the amplitude of
the chaotic motion that is present near the resonance is smaller
than, or at most comparable with,
the amplitude of resonant motion. This is illustrated for the
$(2:1)$ resonance in figures~4 and~5 for the incident right and left
circularly polarized (RCP and LCP) waves, respectively. 
It is clear from these results that there is a high-frequency component
with dominant frequency $\approx \Omega$
that is at least partly chaotic and is superposed on an {\em average}
regular motion. 
The net amplitude of this ``fast'' component can be crudely
estimated to be $\delta L/L\approx 0.005$ and
$\delta G\approx 2\,\delta L$ regardless of polarization; this is in
rough agreement with figures~4 and~5.
For the regular motion
$\Delta_0/2=\Phi_0\pm\tilde{g}_0$; hence, for the RCP case
$\Delta_0/2=-(\pi/6+3^{1/2}/4)$ and
$K^2_+(0.5)\approx 0.923$ so that using equation~\eqref{eq:48} we find
$\delta L\approx 0.094$ in good agreement with the numerical results
in figure~4 when the small-amplitude chaotic signal is ignored. Similarly,
for the LCP case 
$\Delta_0/2=5\pi/6-3^{1/2}/4$,
$K^2_-(0.5)\approx -0.009$, and using equation~\eqref{eq:49} 
we find $\delta L\approx 0.006$ for the amplitude
of regular oscillations as in figure~5. 
It is interesting to note that the amplitude of the ``chaotic'' signal
in $L$ is about the same for RCP and LCP waves. However, the
response of the orbit with $p_\theta>0$ to LCP radiation is smaller
by an order of magnitude than its response to RCP radiation; hence,
the ``chaotic'' signal in figure~5 is comparable in amplitude to the
regular signal that is consistent with the first-order averaged equations.
Thus the chaos that appears in figures~4 and~5 has small
amplitude; in fact, to encounter large-amplitude chaos as in
figures~2 and~3, it seems from our numerical work that the strength
of the perturbation must be larger by at least an order of magnitude.
The variation of the
angular momentum $G$ is equal to the variation of $L$ for $m=2$ and in (out of)
phase with it for RCP(LCP) waves in accordance with equation~\eqref{40} and
in agreement with the numerical data of figures~4 and~5.
Furthermore, it is clear from equation~\eqref{eq:39} that the angle
$\Delta$ satisfies the standard equation for a nonlinear
pendulum to first order in 
$\epsilon^{1/2}$ and the frequency associated with the
small-amplitude oscillations of this pendulum is $\tilde \omega$ given
by
\begin{equation}\label{eq:50}
\tilde{\omega}^2=\frac{3}{2}\epsilon m\Omega^2|\alpha K^m_\pm(\tilde{e}_0)|.
\end{equation}
For the RCP case depicted in figure~4,
we find that $2\pi/\tilde{\omega}\approx 65$. The
pendulum involved here is nonlinear
with amplitude $\Delta_0(\mbox{RCP})$; therefore, the
relevant period of oscillation is longer by about 29\% resulting
in a predicted period of $\approx 84$  in agreement with the
period of regular oscillations depicted in figure~4. Similarly,
for the LCP case depicted in figure~5, we find from equation~\eqref{eq:50}
that the period of small-amplitude  oscillations is $\approx 670$. 
The period of regular oscillations can again be calculated for the 
nonlinear pendulum with amplitude $\Delta_0(\mbox{LCP})$, and it turns
out to be longer by about 10\% in this case due to the nonlinearity. Thus
the predicted period is $\approx 739$, which agrees
with the numerical results based on the
exact system given in figure~5.
\begin{figure}[p]
\epsfxsize=300pt
\centerline{\epsffile{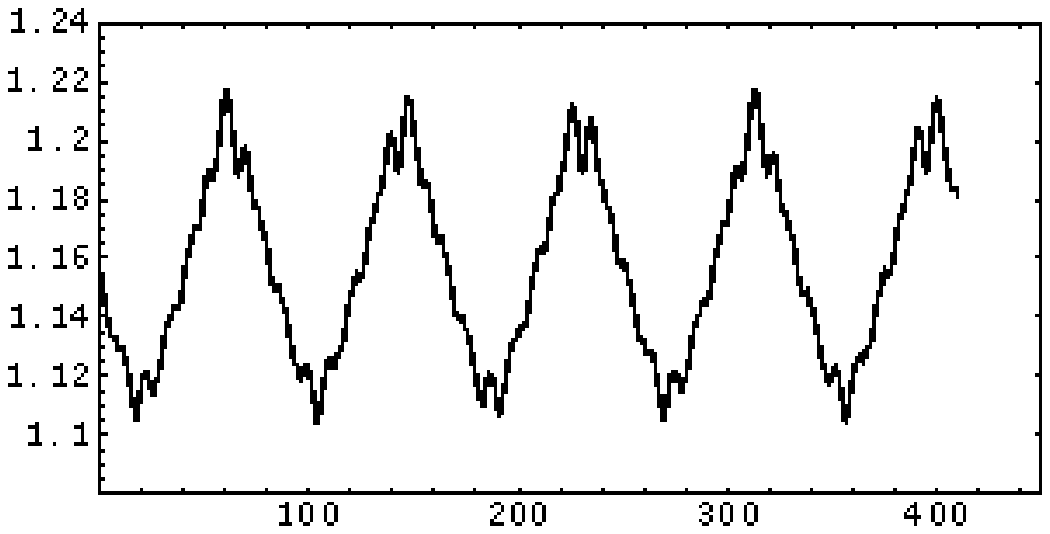}}
\epsfxsize=300pt
\centerline{\epsffile{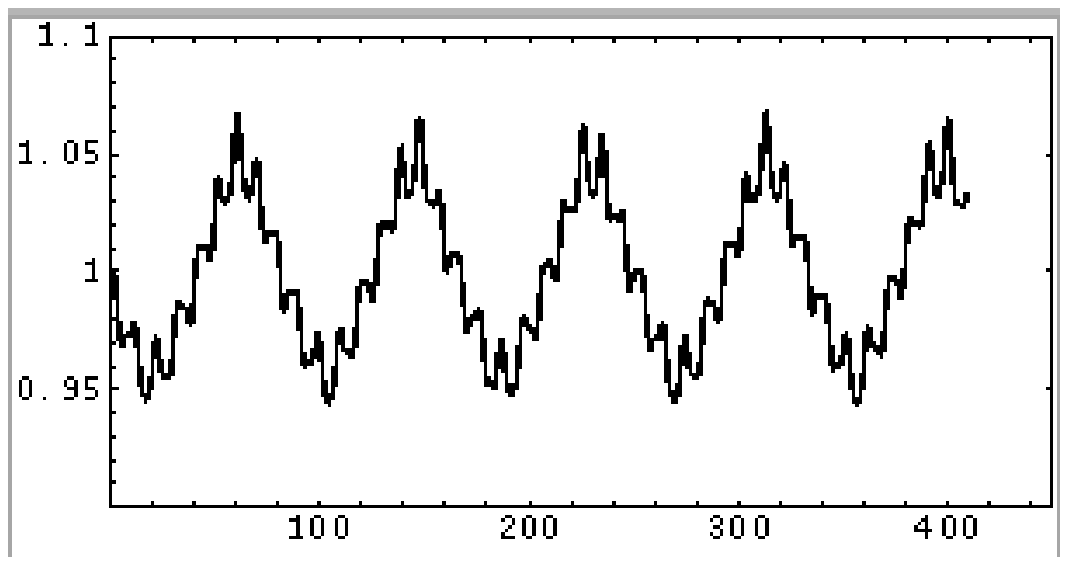}}
\epsfxsize=300pt
\centerline{\epsffile{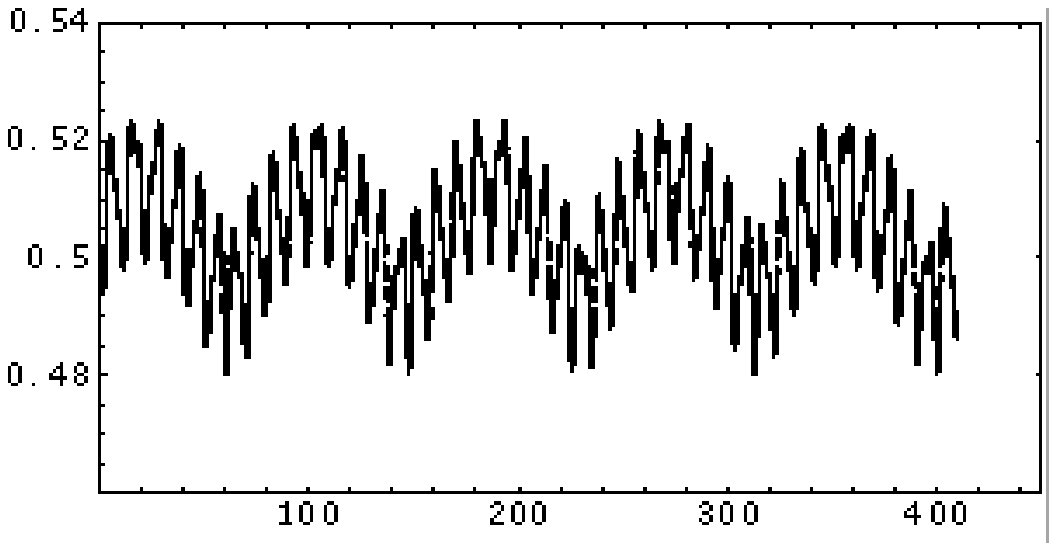}}
\caption[]{The behavior of relative orbit in a binary
system with initial conditions 
$(p_r, p_\theta,r,\theta)=(0.5, 1,1,0)$
in $(2:1)$ resonance with a normally incident {\em right
circularly polarized} (RCP) gravitational wave.
Here $k_0=1$, $\epsilon\alpha=0.002$ and $\Omega=2(3/4)^{3/2}\approx 1.299$. 
The top panel depicts $L$ versus time, the middle panel depicts
the angular momentum $G$ versus time and the bottom panel depicts the
eccentricity $e$ versus time.
}
\end{figure}
\begin{figure}[p]
\epsfxsize=250pt
\centerline{\epsffile{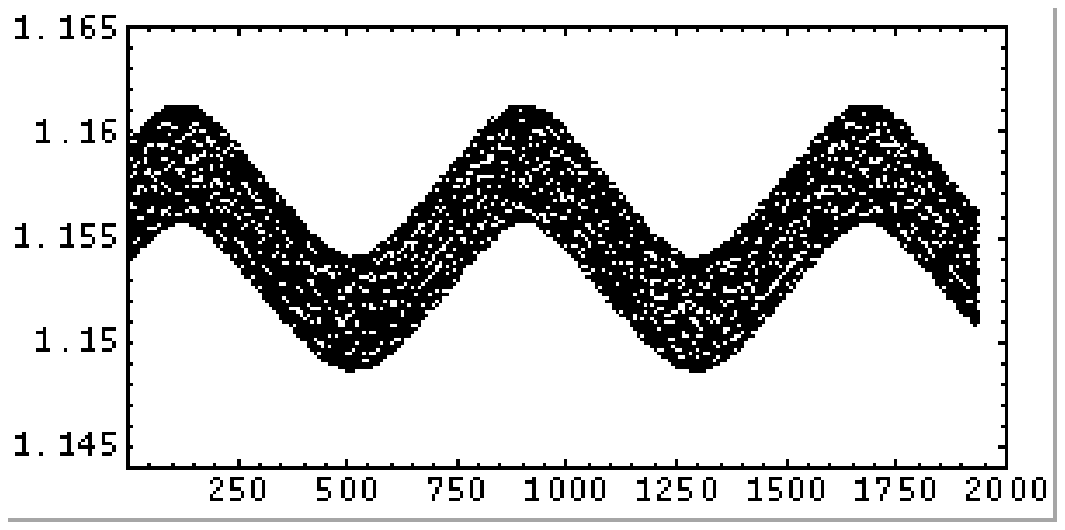}}
\epsfxsize=300pt
\centerline{\epsffile{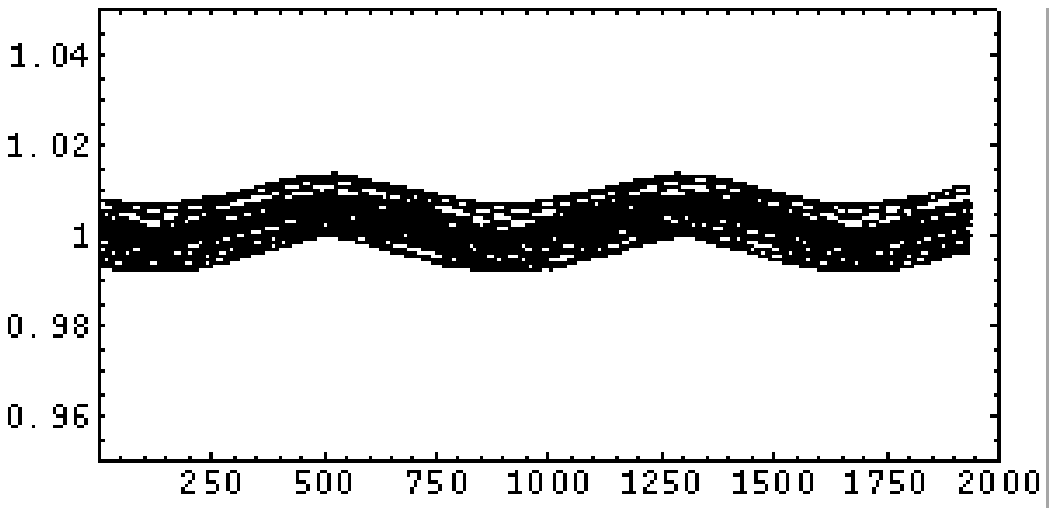}}
\epsfxsize=300pt
\centerline{\epsffile{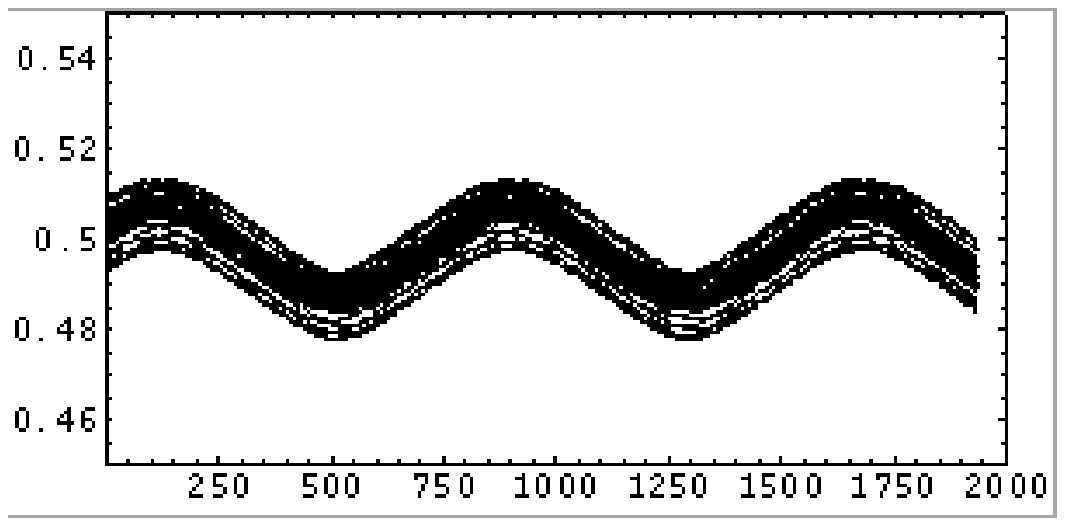}} 
\caption[]{Same as in figure~4, except that the incident radiation
is left circularly polarized (LCP).
}
\end{figure}

Large-amplitude chaos near primary resonances can still occur in
our system if the Chirikov resonance-overlap condition holds. For instance,
numerical experiments with the initial orbit under consideration here
indicate the presence of large-scale chaos if the incident RCP wave
has frequency $\Omega=\pi/2$. To explain the appearance of large-scale
chaos in terms of the Chirikov criterion, let us note that a primary
$(m:1)$ resonance would correspond to a Delaunay variable
$\tilde L_m$ in this case given by 
$\tilde L_m=(2m/\pi)^{1/3}$. 
Thus the initial Delaunay variable $\tilde L$ of the orbit in our example,
$\tilde L_0=(4/3)^{1/3}\approx 1.15$, falls between those of 
$(2:1)$ and $(3:1)$ resonances since $\tilde L_2\approx 1.08$ and
$\tilde L_3\approx 1.24$. The Chirikov criterion for large-amplitude
chaos~\cite{c1,c2} can be written in this case as
\[
\epsilon_{\mbox{ch}}^{1/2}=
\frac{\tilde L_{m+1}-\tilde L_{m}}
{[(2m/3) \tilde L_{m}^2 |\alpha K_+^m(\tilde e_0)|]^{1/2}+
[(2(m+1)/3) \tilde L_{m+1}^2 |\alpha K_+^{m+1}(\tilde e_0)|]^{1/2}},
\]
using equations~\eqref{eq:48} and~\eqref{eq:49}.
{}From $K_+^2(0.5)\approx 0.92$ and $K_+^3(0.5)\approx 0.76$, 
we find that for $m=2$,
\[\epsilon_{\mbox{ch}}\approx 1.65\times 10^{-3}.\]
Our numerical computations suggest, as is expected, that large-scale
chaos occurs for a somewhat smaller choice of the perturbation
parameter,  certainly for $\epsilon=10^{-3}$. 
At much smaller $\epsilon$, for example $\epsilon=10^{-5}$,
the action $L$ still exhibits large-amplitude 
chaotic effects, but seems to be bounded
by KAM tori.
However, for 
$\epsilon\gtrsim\epsilon_{\mbox{ch}}$ 
it appears that the action $L$ 
is no longer confined
by KAM tori. 
This is illustrated in figure~6, where the nature of the binary
orbit is studied for $\Omega=\pi/2$ and $\epsilon=10^{-3}$. At
this Chirikov threshold, all KAM tori are expected to disappear
and gravitational ionization might take place. It should be noted that in
our previous work~\cite{cmr4,6}, the ionization process was associated
with Arnold diffusion. Here, however, gravitational ionization is due
to the fact that $\epsilon>\epsilon_{\mbox{kam}}$.
In figure~6, the perturbed orbit is displayed by plotting
$Y=r\sin\Theta$ versus $X=r\cos\Theta$. In the absence of the
perturbation, $r$ varies from
$\tilde a(1-\tilde e)=2/3$ to $\tilde a(1+\tilde e)=2$ along
the initial elliptical orbit; however,
$r$ can reach relatively large values after the RCP perturbation
is turned on. In the experiment depicted
in figure~6, for instance, the orbit stays generally close
to the initial osculating ellipse for about $100000$ time units,
but large-scale deviations set in eventually such that
$r$  can have values as large as
$\approx 135$ over an interval containing approximately $450000$
time units.   

As in our previous work on gravitational ionization~\cite{5,cmr4,6},
we find that the binary system is not completely destroyed in the
process of gravitational ionization; that is,
one member of the binary does not go away
once and for all. Rather, the system alternates between
dissociation and recombination as in figure~6. This is essentially
due to the oscillatory character of the energy exchange between
the wave and the relative orbit.
In a realistic astrophysical environment, however, new
phenomena might lead to the complete ionization of the binary system
once one member ventures sufficiently far from the other one.
 
Finally, let us assume that the system~\eqref{36} is off resonance,
i.e. it is far from the primary $(m:1)$ resonance manifold. Then we
expect that the motion would involve small amplitude oscillations of
frequency $\approx\Omega$ and amplitude $\sim \epsilon$ near the initial
osculating ellipse. We have explicitly verified this point for the example
given above but with $\Omega=\pi/10$.
\begin{figure}[p]
\epsfxsize=280pt
\centerline{\hspace{-.3in}\epsffile{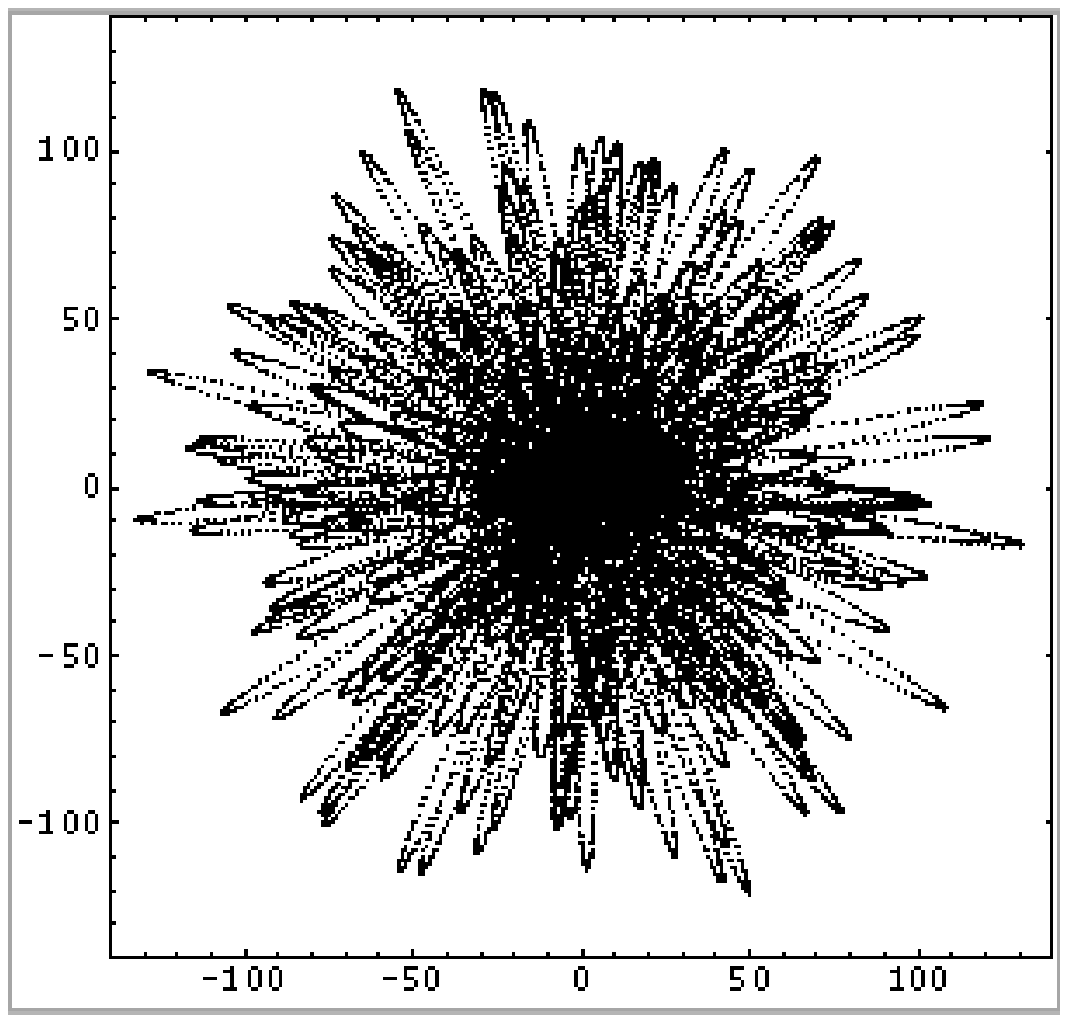}}
\epsfxsize=280pt
\centerline{\epsffile{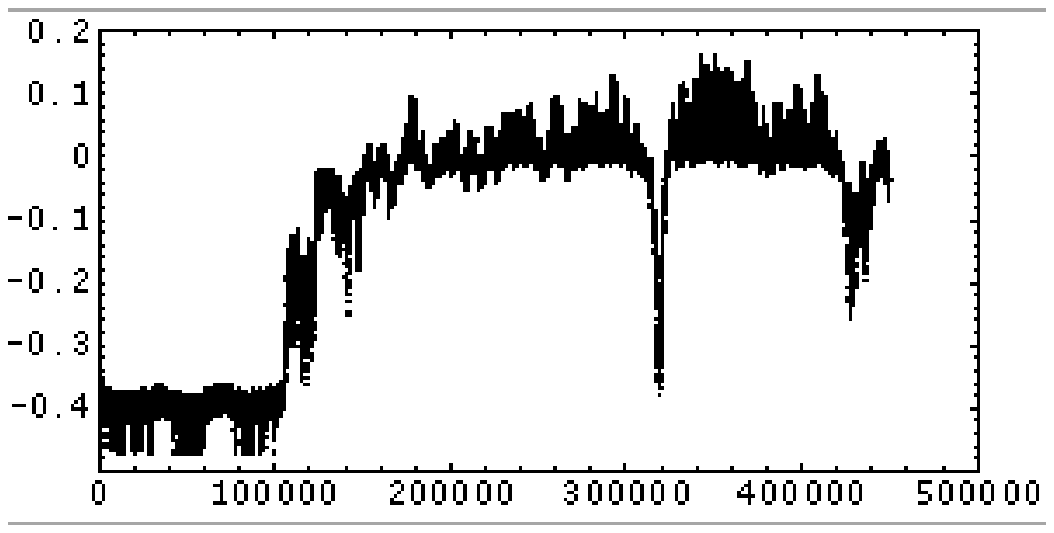}}
\epsfxsize=280pt
\centerline{\epsffile{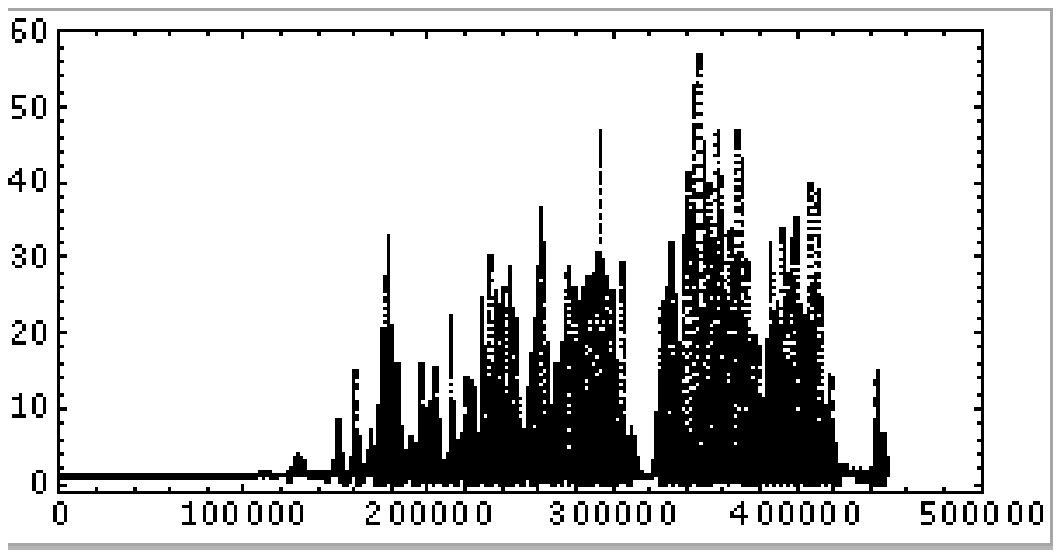}}
\caption[]{Plot of the perturbed 
orbit exhibiting large-amplitude chaos. 
The initial conditions are $(p_r,p_\theta,r,\theta)=(0.5,1, 1,0)$
with $k_0=1$, and the parameters for a normally incident RCP wave
are given by $\epsilon=10^{-3}$, $\alpha=2$ and
$\Omega=\pi/2$. 
The top panel depicts a chaotic rosette corresponding to
the relative orbit of the binary system over
approximately $450000$ time units, the middle
panel depicts the energy $\calE=E$ given by equation~\eqref{eq:ener} 
versus time and the bottom panel depicts the orbital angular momentum 
$p_\theta=G$ versus time. 
}
\end{figure}
\section{Discussion}\label{sec:vii}
The KAM analysis of Hamiltonian chaos is explicitly illustrated
in this paper for the interaction of normally incident
circularly polarized gravitational radiation with a Keplerian
binary system. In the rotating Hill coordinates, the
gravitational wave stands completely still and hence the Hill
system becomes autonomous. The existence of chaos in this system
is demonstrated numerically. On the analytic side, there is no
proof at present that the standard approach to Hamiltonian chaos --- i.e.
the Poincar\'e-Melnikov theory --- is applicable to the Hill system.
Nevertheless, the Melnikov function has been useful in this regard in
certain analogous circumstances~\cite{ds,eks}. Therefore, we compute
the Melnikov function for the Hill system and show that it has the requisite
properties for chaos. The astrophysical implications of these results
are explored; in particular, we employ the partially averaged
equations to describe the nature of the relative
orbit when the binary is at resonance with the incident wave.
The possibility of gravitational ionization in the Hill system
is discussed in connection with large-scale chaotic orbital motions
brought about by incident radiation with an amplitude
$\epsilon\gtrsim\epsilon_{\mbox{ch}}$, where $\epsilon_{\mbox{ch}}$ corresponds to Chirikov's
resonance-overlap criterion.

\appendix
\section{Some Properties of 
    $K^{\lowercase {m}}_\pm(\lowercase{e})$}\label{appen:a}
The quantities 
\[K^m_\pm(e):=\frac{1}{2} m(A_m\pm B_m), \qquad m=1,2,3,\ldots, \]
can be expressed in terms of Bessel functions via equations~\eqref{eq14}
and~\eqref{eq15}.
Using standard expressions for the Bessel functions, we find that for
$e\ll 1$, 
\begin{equation}\label{a:1}
K^m_+(e)=\frac{m^{m-2}}{2^{m-3} (m-2)!}e^{m-2}
        -\frac{(m^2+4m-2) m^{m-1}}{2^{m-1} m!}e^m+O(e^{m+2}).
\end{equation}
Thus, 
$K_+^1(e)=-3 e+O(e^3)$, $K_+^2(e)=2-5 e^2+O(e^4)$,
$K_+^3(e)=3e+O(e^3)$, etc., for $e\to 0$, so that
$K_+^m(0)=2\delta_{m2}$. Moreover, for $e\ll 1$, 
\begin{equation}\label{a:2}
K^m_-(e)=-\frac{(5m+2)m^{m-1}}{2^{m+1} (m+2)!}e^{m+2}+O(e^{m+4});
\end{equation} 
hence, $K_-^m(0)=0$.
It follows that $K^m_-(e)<0$ on the interval $0<e<1$, since
we have already shown in~\cite{5} that $K^m_-(e)$ is monotonically decreasing
on this interval (see p.\ 107 of~\cite{5}).

We are also interested in the function
$K^m_{\pm e}=dK^m_\pm/de$. It is simple to see from~\eqref{a:1} that
$K^1_{+e}(0)=-3$,
$K^3_{+e}(0)=3$, and for all other values of $m\ne 1,3$, 
$K^m_{+e}(0)=0$.
Similarly, it follows from~\eqref{a:2}  that
$K^m_{-e}(0)=0$.

For $e\to 1$, one can use the definition of $K^m_{\pm}$ to find
\begin{equation}\label{a:3}
K^m_\pm(1)=-\frac{2}{m} J_m(m).
\end{equation} 
Here $J_1(1)\approx 0.44$,  $J_2(2)\approx 0.35$, etc., and
for $m\gg 1$
\begin{equation}\label{a:4}
K^m_\pm(1)\sim -\frac{(4/3)^{2/3}}{\Gamma(2/3)} m^{-4/3}.
\end{equation} 
It is also interesting to determine the behavior of 
$K^m_{\pm e}$  as $e\to 1$. We find that
$K^m_{\pm e}(1)$ diverges as
\begin{equation}\label{a:5}
K^m_{\pm e}(e){{} \atop {\displaystyle{\sim}\atop{e\to 1}}}
\pm(1-e^2)^{-1/2}\;\frac{4}{m}J_m'(m),
\end{equation}  
which agrees with the results presented in figure~1.

\section{Calculation of $M(\xi)$}\label{appen:b}
Let us consider the solution $\tau
\mapsto (J_0(\tau ),\sigma _0(\tau ))$ for
the unperturbed heteroclinic orbit of the system~\eqref{eq:un}. 
Starting with equation~\eqref{29}, we note that
the function $\tau \mapsto \sin \sigma_0 (\tau)$
is odd,  so only the odd terms in the
expression for $f_1$ contribute to the
Melnikov integral. Likewise, the function
$\tau \mapsto J_0^2(\tau )$ is even, so only
the even terms of
$\calC_{\ell L}+(2/m)\calC_{\ell G}$ 
and $\partial\Gamma/\partial \sigma$
contribute. To compute $\beta$ in the formula for $f_1$, 
let us consider 
$\calC$ in the form 
\begin{eqnarray*}
\calC (L,G,\ell ,g) &=&\frac{5}{2}
\frac{L^4}{k_0^2}\Big(
1-\frac{G^2}{L^2}\Big) \cos 2g\\ &&\quad +
\frac{L^4}{k_0^2}\sum^\infty_{\nu
=1}\frac{1}{\nu }\Big( K^\nu_+\cos (2g+\nu
\ell )+K^\nu_-\cos (2g-\nu \ell )\Big),
\end{eqnarray*} 
and determine its partial derivatives with respect
to the action variables
\begin{eqnarray*}
\calC_L(L,G,\ell ,g)&=&
\frac{5L}{k_0^2} (2L^2-G^2)\cos 2g\\ &&\quad{}
+4\frac{L^3}{k_0^2}
\sum^\infty_{\nu =1} \frac{1}{\nu} \Big(K^\nu
_+\cos (2g+\nu \ell )
   +K^\nu_-\cos (2g-\nu \ell )\Big)\\ &&\quad{}
+\frac{L^4}{k_0^2}
\sum^\infty_{\nu =1}\frac{1}{\nu} \Big(K^\nu
_{+L}\cos (2g+\nu \ell ) +K^\nu_{-L}\cos
(2g-\nu \ell )\Big)
\end{eqnarray*} and
\begin{eqnarray*}
\calC_G (L,G,\ell, g)&=&
-5\frac{L^2G}{k_0^2}\cos 2g\\  &&\quad{}
+\frac{L^4}{k_0^2}\sum^\infty_{\nu
=1}\frac{1}{\nu }
  \Big(K_{+G}^\nu \cos (2g+\nu \ell )+K^\nu_{-G}
\cos (2g-\nu \ell)\Big),
\end{eqnarray*}  where $K^\nu_{+L}=\partial
K^\nu_{+}/\partial L$, etc. Also, let us
note that here $\ell=(\sigma_0+\pi-2g)/m$ and 
$g=\tau/(6 \mu)$. 
Of course, we must replace $\tau $ by
$\tau +\xi$ in the Melnikov integrand wherever
there is {\em explicit} dependence upon time in $f_1$ and $g_1$. 
For
notational convenience, let us define
\begin{eqnarray*}
\zeta_\pm & =& \frac{\nu}{m}\sigma_0 (\tau )
\pm\frac{\tau}{3\mu }\Big( 1\mp\frac{\nu
}{m}\Big), \\ s_\pm &=& \pm\frac{\xi
}{3\mu }\Big( 1\mp\frac{\nu }{m}\Big)
+\frac{\nu \pi}{m}. 
\end{eqnarray*}  
We then have an expression for $f_1=\Gamma-(m/6)\beta$ using
equation~\eqref{eq:27},
\begin{eqnarray*} f_1 & =& \frac{1}{2}\alpha m
\sum^\infty_{\stackrel{\nu=1}{\scriptscriptstyle
\nu\neq m}} K^\nu _+\frac{\cos
(\zeta_++s_+)}{1-\frac{\nu }{m}}
-\frac{1}{2}\alpha m 
\sum_{\nu =1}^\infty K^\nu_-\frac{\cos
(\zeta_-+s_-)}{1+\frac{\nu}{m}}\\ &&\quad{}
-2J_0^2 -\frac{5}{3}\alpha m^2 
\Big(1-\frac{1}{m}\frac{G_0}{L_0}-\frac{1}{2}
\frac{G_0^2}{L_0^2}\Big)
\cos
\Big(\frac{\tau}{3\mu}+\frac{\xi}{3\mu}\Big)\\ 
&&\quad {}- \frac{2}{3}\alpha m^2
\sum^\infty_{\nu =1}\frac{1}{\nu}
\Big(K^\nu_+\cos (\zeta_++s_+)+K^\nu_-\cos
(\zeta_-+s_-)\Big)\\ &&\quad{} -\frac{1}{6}\alpha
m^2 L_0\sum^\infty_{\nu =1}\frac{1}{\nu}
\Big(K^\nu_{+L} \cos(\zeta_++s_+)+K^\nu_{-L}\cos
(\zeta_-+s_-)\Big)\\ &&\quad {}- \frac{1}{3}\alpha
m L_0 
\sum^\infty_{\nu=1}\frac{1}{\nu}\Big(K_{+G}\cos
(\zeta_++s_+)
  +K_{-G}\cos (\zeta_-+s_-)\Big).
\end{eqnarray*}  Moreover, the odd part of
$f_1$ is given by
\begin{eqnarray*} f_1^{\,\mbox{odd}} &
=&-\frac{1}{2}\alpha m
\sum^\infty_{\stackrel{\nu=1}{\scriptscriptstyle
\nu\neq m}} K^\nu_+\frac{\sin \zeta_+\sin
s_+}{1-\frac{\nu }{m}} 
  +\frac{1}{2}\alpha m\sum^\infty_{\nu
=1}K^\nu_-
\frac{\sin \zeta_-\sin s_-}{1+\frac{\nu
}{m}}\\  
&&\quad{} +\frac{5}{3}\alpha m^2
\Big(1-\frac{1}{m}\frac{G_0}{L_0}-\frac{1}{2}\frac{G^2_0}{L^2_0}\Big)
\sin \big(\frac{\tau }{3\mu}\big)\sin\big( \frac{\xi}{3\mu}\big)\\  
&&\quad{}
+\frac{2}{3}\alpha m^2
\sum^\infty_{\nu =1}
\frac{1}{\nu} (K^\nu_+\sin \zeta_+\sin s_+
+K^\nu_-\sin \zeta_-\sin s_-)\\  &&\quad{}
+\frac{1}{6}\alpha m^2 L_0
\sum^\infty_{\nu =1}
\frac{1}{\nu}  (K^\nu_{+L}\sin \zeta_+\sin
s_++K^\nu_{-L}\sin \zeta_-\sin s_-)\\ 
&&\quad{}+\frac{1}{3}\alpha m L_0
\sum^\infty_{\nu =1}\frac{1}{\nu} 
(K^\nu_{+G} \sin \zeta_+\sin
s_++K^\nu_{-G}\sin \zeta_-\sin s_-).
\end{eqnarray*}  
On the other hand,  only
the {\em even terms} of
$\calC_{\ell L} +(2/m)\calC_{\ell G}$ 
contribute to $M$; to compute these
terms, we differentiate the expressions for $\calC_L$ and 
$\calC_G$ with respect to $\ell$, 
\begin{eqnarray*}
\calC_{\ell L} & =&
\frac{4L^3}{k_0^2}\sum^\infty_{\nu =1}
\Big(-K^\nu_+\sin (2g+\nu \ell ) +K^\nu
_-\sin (2g-\nu \ell )\Big)\\  &&\quad{} +
\frac{L^4}{k_0^2}\sum^\infty_{\nu =1}
\Big(-K^\nu_{+L} \sin (2g+\nu
\ell)+K^\nu_{-L}\sin (2g-\nu \ell)\Big),\\
\calC_{\ell G} &
=&\frac{L^4}{k_0^2}\sum^\infty_{\nu =1}
\Big(-K^\nu _{+G}\sin (2g+\nu
\ell)+K^\nu_{-G} \sin (2g-\nu \ell)\Big).
\end{eqnarray*}  Thus, we find that at
resonance
\begin{eqnarray*}
\calC_{\ell L} +\frac{2}{m}\calC_{\ell G} &
=& -\frac{4m}{\Omega} \sum^\infty_{\nu =1}
\Big(K^\nu _+\sin (\zeta_++s_+)+K^\nu _-\sin
(\zeta_-+s_-)\Big)\\  &&\quad{} 
-\frac{mL_0}{\Omega}\sum^\infty_{\nu =1}
\Big(K^\nu_{+L}\sin (\zeta_++s_+)+K^\nu_{-L}\sin
(\zeta_-+s_-)\Big)\\ &&\quad{}
-\frac{2L_0}{\Omega}
\sum^\infty_{\nu =1} \Big(K_{+G}^\nu \sin
(\zeta_++s_+)+K^\nu_{-G}\sin(\zeta_-+s_-)\Big)
\end{eqnarray*}  and the even terms are
\begin{eqnarray*}
\Big(\calC _{\ell L}+\frac{2}{m}\calC_{\ell
G}\Big)^{\mbox{even}} & =&
-\frac{4m}{\Omega}\sum^\infty_{\nu =1}
(K^\nu _+\cos \zeta_+\sin s_++K^\nu_-\cos
\zeta_-\sin s_-)\\  &&\quad{}
-\frac{mL_0}{\Omega}\sum^\infty_{\nu =1}
(K^\nu_{+L}\cos \zeta_+\sin s_+
+K^\nu_{-L}\cos \zeta_-\sin s_-)\\  &&\quad{}
-\frac{2L_0}{\Omega}\sum^\infty_{\nu =1}
(K^\nu_{+G}\cos \zeta_+\sin s_+
+K^\nu_{-G}\cos \zeta_-\sin s_-).
\end{eqnarray*}

Finally, we must compute $-J_0^2\;\partial\Gamma/\partial\sigma$, where
\[
\frac{\partial \Gamma}{\partial\sigma}=
-\frac{1}{2}\alpha
\left[\,
\sum^\infty_{\stackrel{\nu=1}{\scriptscriptstyle
\nu\neq m}}
\nu K^\nu_+
\frac{\sin(\zeta_++s_+)}{1-\frac{\nu}{m}}
-\sum^\infty_{\nu=1}
\nu K^\nu_-
\frac{\sin(\zeta_-+s_-)}{1+\frac{\nu}{m}}
\right]
\] using equation~\eqref{eq:27}.  Only the
even terms in $\partial
\Gamma/\partial\sigma$ would contribute to
the Melnikov integral, hence
\[
\Big(\frac{\partial
\Gamma}{\partial\sigma}\Big)^{\mbox{even}}= -\frac{1}{2}\alpha
\left[\,
\sum^\infty_{\stackrel{\nu=1}{\scriptscriptstyle
\nu\neq m}}
\nu K^\nu_+ \frac{\cos\zeta_+\sin
s_+}{1-\frac{\nu}{m}} -\sum^\infty_{\nu=1}
\nu K^\nu_- \frac{\cos\zeta_-\sin
s_-}{1+\frac{\nu}{m}}
\right].
\]

We now have
\begin{eqnarray*} M(\xi)&=& -\frac{1}{6}
\alpha mK^m_+\int^\infty_{-\infty}\sin
\sigma_0 (\tau ) f_1^{\mbox{odd}}\, d\tau 
+\frac{1}{6}\alpha \Omega
\int^\infty_{-\infty} J_0^2(\tau )
\Big(\calC_{\ell L}+\frac{2}{m} \calC_{\ell
G}\Big)^{\mbox{even}}\,d\tau \\ &&\quad{}
-\int^\infty_{-\infty}
J_0^2(\tau)\Big(
\frac{\partial\Gamma}{\partial\sigma_0}\Big)^{\mbox{even}}\,d\tau.
\end{eqnarray*}  Let us define
\begin{eqnarray*}
 I^0&:=&-\frac{1}{6} \alpha m K^m_+
\int^\infty_{-\infty}\sin \sigma_0
(\tau)\sin\Big(\frac{\tau }{3\mu
}\Big)\,d\tau ,\\ 
I^\pm_\nu  &:=&-\frac{1}{6}
\alpha m K^m_+
\int^\infty_{-\infty}\sin \sigma_0 (\tau )\sin
\zeta_\pm\,d\tau, \\  
\calS^0 &:=& \sin \frac{\xi }{3\mu }, \quad 
\calS^\pm_\nu :=\sin s_\pm,\\
J_\nu^\pm  &: =&\int^\infty_{-\infty} 
J_0^2(\tau)\cos \zeta_\pm\,d\tau ,
\end{eqnarray*}  
and note that
$$I_\nu^+=-I^-_{-\nu},\quad
J^+_\nu=J^-_{-\nu},
  \quad 
\calS^+_\nu =-\calS^-_{-\nu}.$$ It follows
that
\begin{eqnarray}
\nonumber M(\xi ) & =& -\frac{1}{2}\alpha m
\sum^\infty_{\stackrel{\nu=1}{\scriptscriptstyle
\nu\neq m}}
\frac{K^\nu _+}{1-\frac{\nu}{m}}I^+_\nu
\calS^+_\nu  +\frac{1}{2}\alpha m
\sum^\infty_{\nu =1}
\frac{K^\nu_-}{1+\frac{\nu}{m}} I^-_\nu
\calS^-_\nu\\ 
\nonumber &&\quad{} +\frac{5}{3}\alpha m^2 
\Big(1-\frac{1}{m}\frac{G_0}{L_0}-\frac{1}{2}
\frac{G^2_0}{L^2_0}\Big) I^0\calS^0 \\
\nonumber &&\qquad {}+\frac{2}{3}\alpha m^2
\sum^\infty_{\nu =1}\frac{1}{\nu}
(K^\nu_+I^+_\nu\calS^+_\nu +K_-^\nu I^-_\nu
\calS^-_\nu)\\ 
\nonumber &&\qquad{} +\frac{1}{6}\alpha m^2
L_0 
\sum^\infty_{\nu=1}\frac{1}{\nu}
(K^\nu_{+L}I^+_\nu\calS^+_\nu+K^\nu_{-L}I^-_\nu
\calS^-_\nu)\\ 
\nonumber &&\qquad{} +\frac{1}{3}\alpha m L_0
\sum^\infty_{\nu =1}
\frac{1}{\nu}
(K^\nu_{+G}I_\nu^+\calS_\nu^++K^\nu_{-G}I^{-}_\nu
\calS^-_\nu)\\ 
\nonumber &&\qquad{} -\frac{2}{3}\alpha m
\sum^\infty_{\nu =1} (K^\nu_+ J^+_\nu
\calS^+_\nu +K^\nu_-J^-_\nu\calS^-_\nu)\\ 
\nonumber &&\qquad{}  -\frac{1}{6}\alpha m L_0
\sum^\infty_{\nu=1} (K_{+L}^\nu J^+_\nu
\calS^+_\nu +K^\nu_{-L}J_\nu^-\calS^-_\nu )\\
\nonumber &&\qquad{}  -\frac{1}{3}\alpha L_0
\sum^\infty_{\nu=1}  (K^\nu_{+G}J^+_\nu
\calS^+_\nu +K^\nu_{-G}J^-_\nu \calS^-_\nu)\\
&&\qquad {}+\frac{1}{2}\alpha
\sum^\infty_{\stackrel{\nu=1}{\scriptscriptstyle
\nu\neq m}}
\frac{\nu
K_+^\nu}{1-\frac{\nu}{m}}J^+_\nu\calS^+_\nu
-\frac{1}{2}\alpha
\sum^\infty_{\nu=1} \frac{\nu
K_-^\nu}{1+\frac{\nu}{m}}J^-_\nu\calS^-_\nu.
\label{eq:29}
\end{eqnarray}  Furthermore, let us define 
\begin{eqnarray*} K^\nu_{\pm
G}&=&\frac{\partial e}{\partial G}
\frac{d}{de} K^\nu_{\pm}
                 :=\frac{\partial
e}{\partial G}K^\nu_{\pm e},\\ K^\nu_{\pm
L}&=&\frac{\partial e}{\partial
L}\frac{d}{de} K^\nu_{\pm}
               :=\frac{\partial e}{\partial
L}K^\nu _{\pm e},\\ F_m(e)&:=&L_0\Big(
\frac{\partial e}{\partial L}
         +\frac{2}{m}\frac{\partial
e}{\partial G}\Big)\Big|_{(L_0,G_0)}. 
\end{eqnarray*} 
We recall that 
$e^2=1-G^2/L^2$; hence, 
$$ e\frac{\partial e}{\partial
L}=\frac{G^2}{L^3},
\quad e\frac{\partial e}{\partial G}
=-\frac{G}{L^2},
$$  so that
$$
\frac{\partial e}{\partial L}=\frac{1}{eL}
(1-e^2),\quad
\frac{\partial e}{\partial G}=-\frac{1}{eL}
(1-e^2)^{1/2},
$$  and thus
$$ F_m(e)=\frac{1}{e}
\left[1-e^2-\frac{2}{m}(1-e^2)^{1/2}\right].$$
Moreover, it is useful  to define
$$H_m(e):=1+e^2-\frac{2}{m} (1-e^2)^{1/2},$$ 
so that
$$H_m(e)-eF_m(e)=2e^2.$$  Finally, let us
define
$$P^\pm_\nu :=\frac{1}{\nu
}I^\pm_\nu-\frac{1}{m} J_\nu^\pm.$$
After collecting terms in
equation~\eqref{eq:29} and using the
definitions given above, we have an
expression for the Melnikov function:
\begin{eqnarray} M(\xi ) &=& \frac{5}{6}
\alpha m^2H_m(e)I^0\calS^0\nonumber\\ 
&&\quad -\frac{1}{2} \alpha m
\sum^\infty_{\stackrel{\nu=1}{\scriptscriptstyle
\nu\neq m}}
\frac{\nu
K^\nu_+}{1-\frac{\nu}{m}}P_\nu^+\calS_\nu^+
+\frac{1}{2} \alpha m \sum_{\nu =1}^\infty
\frac{\nu K^\nu_-}{1+\frac{\nu}{m}}
P^-_\nu\calS^-_\nu\nonumber\\  &&\quad
+\frac{2}{3}
\alpha m^2\sum^\infty_{\nu =1} (K_+^\nu
P_\nu ^+\calS^+_\nu +K^\nu_-P^-_\nu
\calS^-_\nu)\nonumber\\  &&\quad +\frac{1}{6}
\alpha m^2F_m(e)\sum^\infty_{\nu
=1}(K_{+e}^\nu P^+_\nu \calS^+_\nu
+K^\nu_{-e}P^-_\nu
\calS^-_\nu).\label{eq31}
\end{eqnarray} 
Let us note that in this
expression $s_+(\nu =m)=\pi$ so that
$\calS^+_m=0$. 
Furthermore, we have from the definition of
$I^\pm_\nu$ and the fact that 
$\delta \sin \sigma_0(\tau )=-dJ_0/d\tau $,
$$I^\pm_\nu
=\int^\infty_{-\infty}-\frac{dJ_0}{d\tau
}\sin \zeta_\pm d\tau
=\int^\infty_{-\infty}J_0\left[\frac{\nu}{m}J_0\pm \frac{1}{3\mu }
\Big( 1\mp \frac{\nu }{m}\Big)\right]\cos \zeta _\pm\, d\tau ,$$
via integration by parts $(J_0\to 0$ as
$\tau \to \pm \infty$). It thus follows that
$P^\pm _\nu =\nu^{-1}I^\pm _\nu -m^{-1}J^\pm
_\nu $ is given by equation~\eqref{31}. It is
interesting to note that in the formula for
$M(\xi )$, $P^+_m=0$; moreover, the quantity
\begin{equation}\label{b:3}
\frac{P^\pm_\nu}{1\mp \nu/m }
  =\pm \frac{1}{3\nu \mu }
\int^\infty_{-\infty}J_0(\tau )\cos \zeta_\pm\,d\tau 
\end{equation}
is well defined even for $\nu =m$.
In fact, we find from equation~\eqref{eq:un} that
\[
\lim_{\nu\to m} \frac{P_\nu^+}{1- \nu/m}=\frac{1}{3 m \mu}
\int_{-\infty}^\infty J_0(\tau)\cos\sigma_0\, d\tau=0.
\]
Using these results in equation~\eqref{eq31},
we recover equation~\eqref{30} for $M(\xi)$. 

The function $M(\xi)$ refers to a resonance between a Keplerian
orbit and an incident circularly polarized wave. Therefore,
$M(\xi)$ depends upon the wave amplitude $\alpha$, the order
of the resonance $m$, the orbital eccentricity at
resonance $e_0$ and the perturbation parameter $\mu$.
The explicit form of $M (\xi)$ involves certain combinations
of the Bessel functions (i.e. $K^\nu_\pm$ and $K^\nu_{\pm e}$) 
 that depend only on the eccentricity and
the integrals in $I^0$ and $P^\pm_\nu$. The complete evaluation
of  $P^\pm_\nu$ is beyond the scope of this work; however, 
 $P^\pm_\nu$ can be computed using contour integration when
$2\nu/m$ is an integer. To see this, let us assume that
$\delta=-(\alpha m/6)K^m_+(e_0)>0$ and choose the principal branch
of $\sigma_0(\tau)$ in the following. Hence the integral in 
 $P^\pm_\nu$ takes the form
\begin{equation}\label{b:4}
\calI_{jw}=2\int_{-\infty}^\infty \sech(x)\cos[j\arcsin(\tanh x)+w x]\,dx,
\end{equation}
where $j=2\nu/m$ and $w=\pm(1\mp\nu/m)/(3\mu\delta^{1/2})$.
For the analytic extension of the integrand to the complex plane, it is
essential to rewrite the integral in the form
\[
\calI_{jw}=\int_{-\infty}^\infty (\sech x)^{j+1}
  [(1+i\sinh x)^j\exp(iwx)+ (1-i\sinh x)^j\exp(-iwx)]\,dx,
\]
that is appropriate for the principal branch of $\sigma_0$ under
consideration here. We note that the integrand is an even function.
Let us now imagine a rectangular contour in the complex $(x,y)$-plane
that is symmetric about the $y$-axis
with the two long (eventually infinite) sides parallel to the
$x$-axis at $y=0$ and $y=2\pi$. The singularities of the integrand within
this contour occur at $i\pi/2$ and $3i\pi/2$. These are poles once
$j$ is a positive integer. In this case, a standard application of
the Cauchy Theorem for this contour results in
\[
[1-\cosh(2\pi w)]\;\calI_{jw}=2\pi i\;[\mbox{residue}\;(i\pi/2)
+\mbox{residue}\;(3i\pi/2)],
\]
where the symmetry of the integrand has been taken into account. The
calculation of the residues for arbitrary $j$ is straightforward, but
will not be attempted here; however, it is possible
to show that $\calI_{jw}$ has the general form
\begin{equation}\label{b:5} 
\calI_{jw}=4\pi w\frac{\exp(-\pi w/2)}{\sinh(\pi w)}W_j(w),
\end{equation}
where $W_j(w)$ is a polynomial in $w$ of degree $j-1$ such that
$jW_j(0)$ vanishes for even $j$ and is equal to $(-1)^{(j-1)/2}$ for odd
$j$. We find that $W_1(w)=1$, $W_2(w)=-w$, 
$W_3(w)=(-1+2 w^2)/3$, etc.  Note that $\calI_{jw}$ is well defined
for $w\to 0$.
Moreover, $I^0$ can also
be computed using the approach outlined above. We find
that
\[
\int_{-\infty}^\infty \sin(w_0 x)\sin\sigma_0(x)\,dx
=\frac{2\pi w_0}{\cosh (\pi w_0/2)}
\]                                
for the principal branch of $\sigma_0$. For $I^0$,
$w_0=1/(3\mu\delta^{1/2})$; hence,
\begin{equation}\label{b:6} 
I^0=\frac{2\pi}{3\mu\cosh[\pi/(6\mu\delta^{1/2})]}.
\end{equation}

{\bf Acknowledgments.} The work of the first author 
was supported by the NSF 
grant DMS-9531811 and the University of Missouri Research Board.

\end{document}